\documentstyle[onecolumn,epsf]{mn}
\newcommand{\B}{{\bmath B}}
\newcommand{\E}{{\bmath E}}
\newcommand{\RS}{r_* }                              
\newcommand{\GJ}{\rho_{\mbox{\tiny GJ}} }           
\newcommand{\eGJ}{\bmath E^{\mbox{\tiny GJ}}}       
\newcommand{\egj}{ E^{\mbox{\tiny GJ}}}             
\newcommand{\pGJ}{\Psi_{\mbox{\tiny GJ}}}           
\newcommand{\xicrit}{\tilde{\xi}_{\mbox{cr}}}       
\newcommand{\xiDL}{\tilde{\xi}}                     

\newcommand{\V}{ {\bmath v}      }
\newcommand{\J}{ {\bmath J}      }
\newcommand{\n}{ {\bmath \nabla} }
\newcommand{\x}{ {\bmath \times} }
\newcommand{\N}{ {\bmath e_r}    } 
\newcommand{\D}{  \Delta         }
\newcommand{\pd}{ \partial       }

\newcommand{\I}{  {\rm i} }
\newcommand{\dd}{ {\rm d} }

\newcommand{\DSt}{\displaystyle}

\newcommand{\be}{\begin{equation}}
\newcommand{\ee}{\end{equation}}

\begin{document}

\title{Magnetosphere of Oscillating Neutron Star. Nonvacuum
Treatment.}
\author[A.N.Timokhin et al.]%
{A.N.~Timokhin,$^{1,2,3}$%
  \thanks{Current address: Max-Planck-Institut f\"ur Kernphysik,
          Saupfercheckweg~1, D-69117 Heidelberg, Germany} 
 G.S.~Bisnovatyi-Kogan,$^2$ 
 H.C.~Spruit$^3$\\
$^1$Sternberg Astronomical Institute, Universitetskij pr. 13,
119899 Moscow, Russia\\
$^2$Space Research Institute, Profsoyuznaya 84/32, 117810 Moscow,
Russia\\
$^3$Max-Planck-Institut f\"ur Astrophysik, Karl-Schwarzschild Str.1,
D-85740 Garching, Germany
}

\maketitle
\begin{abstract}
We generalize a formula for the Goldreich-Julian charge density ($\GJ$),
originally derived for rotating neutron star,
for arbitrary oscillations of a neutron star with
arbitrary magnetic field configuration under assumption
of low current density in the inner parts of the magnetosphere.
As an application we consider
toroidal oscillation of a neutron star with dipole magnetic field
and calculate energy losses. For some 
oscillation modes the longitudinal electric field can not be
canceled  by putting charged particles in the magnetosphere
without a presence of strong electric current
($j \simeq \frac{c}{\omega r}\, \GJ\: c$).
It is shown that the energy losses are strongly affected by
plasma in the magnetosphere, and cannot be
described by vacuum formulas.
\end{abstract}


\section{Introduction}

Oscillations of neutron stars
could give an opportunity to study their internal structure.
Oscillations of a neutron star (NS) may be observable
in pulsars just after a glitch. The energy release during the glitch
is estimated as 
\be
E \sim \frac{1}{2} \frac{ M_{\mbox{\tiny crust}} }{ M_{\mbox{\tiny NS}} } 
                                       I\Omega^2 \frac{\Delta\Omega}{\Omega}\: ,
\ee 
where $I$ is the moment of inertia of the NS, 
$\Omega$ is an angular velocity of rotation, 
$\Delta \Omega$ is a jump of angular velocity, 
$M_{\mbox{NS}}$ is a mass of the NS, $M_{\mbox{crust}}$ is a mass of the crust.
For the Vela pulsar $E \sim 10^{37}$ergs.
There are about 20 glitching pulsars, which had suffered 
a total of 45 glitches \cite{Lyne96}. 
The energy release in each of  these
glitches is similar to that of the Vela pulsar. 
If several percent of this energy goes into excitation of 
oscillations it would be possible to observe them.
Another example how oscillations of NS could be observed
in pulsars is by observations of microstructure in single pulses.
Boriakoff \shortcite{Boriakoff76} proposed that vibrations 
of a neutron star  may cause the periodicity of micropulses observed 
in some pulsars. A difficulty in this theory is that
 no effective mechanism has been proposed
for excitation of oscillations \cite{Hankins96}.
   
Oscillating neutron stars have been proposed as a source for
Galactic Gamma-Ray Bursts (GRBs)
by Pachini \& Ruderman~\shortcite{PR74}  and Tsygan~\shortcite{Ts75}.
This idea is further developed by Blaes et al.~\shortcite{Blaes89}, 
Smith \& Epstein~\shortcite{Epstein93}, 
Fatuzzo \&  Melia~\shortcite{Melia93}.
Oscillation induced 
hard gamma radiation from NS was proposed by
Bisnovatyi-Kogan~\shortcite{BK95} and Cheng \& Ding~\shortcite{Cheng97}
for explanation of the hard delayed emission observed from some GRBs \cite{Hurley94}, 
even if the Gamma-Ray Burst itself is generated 
by a different mechanism \cite{BK75}. 
Recent GRBs observations make
a Galactic origin of GRBs unlikely, but a Galactic model may not
be excluded completely. 
At least for Soft Gamma Repeaters, which are believed to be neutron stars, 
oscillations of star can play an essential role (Duncan 1998). 

Eigenfrequencies and eigenfunctions of
neutron star oscillations have been computed by many
authors \cite{McD88,Carroll86,Strohmayer91}. 
These calculations have shown
that typical periods of neutron star oscillations range from
$0.1$ up to several tens of milliseconds. On the other hand,
observed pulsar radiation in different spectral regions is generated
mostly in their magnetospheres \cite{Truemper98}.
If we are looking for oscillation induced radiation from pulsars
we should investigate processes in the magnetosphere. 
A complete solution for vacuum electromagnetic fields near an
oscillating magnetized neutron star was obtained by 
Muslimov and Tsygan~\shortcite{MT86}. But for typical NS's magnetic field 
strength ($>10^9$G) the electric field arising from star`s oscillation
will be strong enough to pull charged particles from the surface in the 
magnetosphere. Hence, any realistic model of the magnetosphere of
oscillating NS must take into account the presence of charged particles.
This will also affect the electromagnetic energy losses of a pulsating
magnetized star. 
Electromagnetic energy losses of an oscillating NS in vacuum were calculated by 
McDermott at al.~\shortcite{McD84} and Muslimov and Tsygan~\shortcite{MT86}

As a first step in the investigation of magnetosphere of an oscillating
neutron star with nonzero charge density
we consider the inner part of the magnetosphere, assuming 
low current density there.
We generalize a formula
for Goldreich-Julian charge density, derived originally for rotating NS, 
for a NS oscillating in any arbitrary mode.
This allow us to take into account plasma present in the magnetosphere 
of the NS and calculate the energy losses. The plan of the paper
is following. In the first section we introduce some definitions
and give algorithm for calculation of Goldreich-Julian charge density
for arbitrary oscillation mode and magnetic field structure under
the assumption of low current density, and 
discuss the limitations of this assumption.
In the second section we apply this formalism to
toroidal oscillations of a NS with a dipole magnetic field.
We calculate the Goldreich-Julian charge density 
and  energy losses for some oscillation modes and discuss 
factors affecting energy losses of oscillating neutron star.

\section{Goldreich-Julian charge density for small amplitude
oscillations of a neutron star. General formalism.}

\subsection{Basic definitions.}

As it was firstly pointed by Goldreich and Julian \cite{GJ69} a rotating NS can 
not be surrounded by vacuum. The electric field generated by rotation of magnetized
NS will pull charged particles into the magnetosphere -- the electrostatic force near
the NS's surface in vacuum would be much stronger than the gravitational one for
both electrons and ions. 
According to the calculations of Jones~\shortcite{Jones86}, the binding of
charged particles in the crust of a typical pulsar is not strong enough 
to prevent them being pulled into the magnetosphere by the vacuum electric field. 
But even if the binding of charged particles in the cold crust of an old NS can
prevent them escaping, the magnetosphere of a strong magnetized neutron star 
can be filled by charged particles by the mechanism
proposed by Ruderman and Sutherland \shortcite{Ruderman75} for pulsars.

On the NS surface the vacuum electric field generated by rotation or
oscillation has a radial component, whose value is a substantional part
of the full field strength \cite{MT86,GJ69}. To order of magnitude this
field for the rotating star is 
\be
E^{\rm rot} \simeq \frac{\Omega \RS}{c} B \ ,
\ee
where $\Omega$ is the rotational frequency of the NS, $\RS$ is NS radius and $B$ is 
the magnetic field strength near the star. In the case of NS oscillations the
corresponding vacuum electric field is
\be
E^{\rm osc} \simeq \frac{\omega \xi}{c} B \ ,
\ee
where $\omega$ is the oscillation frequency, $\xi$ is the displacement amplitude.
The vacuum electric field strength near the NS oscillating with period $\tau$ will be
of the same order as the field strength generated by rotation of the NS with period
$T_{\rm eff}$:
\be
T_{\rm eff} \simeq \tau \frac{\RS}{\xi}
\: .
\label{t_rot_same_as_osc}
\ee
For typical oscillation parameters \cite{McD88} the strength of the vacuum electric
field of the oscillating NS will be the same as that generated 
by a $\sim 1$~sec pulsar.
Hence, the magnetosphere of an oscillating, even non-rotating, NS should be filled by
charged particles. 
In the strong magnetic field of the NS ($B>10^9$G) synchrotron energy losses of
charged  
particles in the magnetosphere are very high. They lose their perpendicular
momentum (if any) very rapidly and occupy the first Landau level, i.e. they can
move only along magnetic field lines. 

In the presence of strong longitudinal (parallel to $\B$)
electric field $E_{\|}$ charged particles are accelerated to high
energies and their curvature radiation in strong magnetic field produces
an electron-positron pair cascade \cite{Sturrock71,Harding82}.
The  particles produced in the cascade cancel the accelerating electric field.
Because of this there should be a regular force-free ($E_{\|} \ll B$) configuration of
the electromagnetic field, at least in the inner parts of the magnetosphere, where
the NS magnetic field is strong enough to allow single photon pair creation:
\be
\eGJ \cdot \B = 0 
\: .
\label{force-free} 
\ee
Let us call the electric field $\eGJ$ the Goldreich-Julian (GJ)
electric field. We introduce the generalized Goldreich-Julian charge density as
the charge density in the magnetosphere corresponding to the GJ electric field
$\eGJ$:
\be
\GJ = \frac{1}{4\pi} \n \cdot \eGJ\: .
\label{RhoGJ}
\ee
It follows from the above discussion, that the charge density in the inner parts
of the magnetosphere of an oscillating NS should be approximately
equal to the GJ charge density. 

In the following we are looking for the
GJ electric field and the GJ charge density, because they should adequately describe
the electric field and the charge density in the inner parts of the NS magnetosphere. 
Knowledge of them will allow us to calculate electromagnetic 
energy losses of an oscillating NS.

\subsection{Basic equations. Low current density approximation.
\label{small_current_approx_section}}

We assume the neutron star to be a magnetized conducting sphere
of a radius $\RS$. We are interested only in the  pulsation modes with
non vanishing amplitude  $\xi$ on the surface. 
Consider a region close to the NS (near zone),
at distances from the NS surface  smaller than the wave length 
$\lambda = 2 \pi c/\omega$, where $\omega$ is the oscillation 
frequency and $c$  is the speed of light.
In the near zone $r < \lambda $  one can 
neglect the displacement current term.
Maxwell`s equations for the electric and magnetic fields in the near zone are:
\begin{eqnarray}
\n \cdot \E   & = & 4 \pi \rho
\: ,
\label{max1}\\
\n \times \E  & = & - \frac{1}{c} \partial_{t} \B
\: ,
\label{max2}\\
\n \cdot \B   & = & 0
\: ,
\label{max3}\\
\n \times \B  & = & \frac{4\pi}{c} {\bmath j}
\: ,
\label{max04}
\end{eqnarray}
where $\B$ and $\E$ are the magnetic and electric fields, $\rho$ and ${\bmath j}$ are
the charge and current density. We use in the notation for
partial derivatives:  
$\pd_\alpha \equiv \pd/\pd \alpha$.
On the unperturbed surface of the NS $\E$ and $\B$ must satisfy the
following boundary conditions:
\begin{eqnarray}
B_r(\RS)              & = &   B_{0r}
\: ,
\label{bBC1}\\
B_{\theta, \phi}(\RS) & = &   B_{0\theta, 0\phi} \pm  \frac{4 \pi}{c}\J_{\phi, \theta}
\: ,
\label{bBC2} \\
E_{\theta, \phi}(\RS) & = &  -\frac{1}{c}(\V \x \B_0)_{\theta, \phi}  
\: ,
\label{eBC1}\\
E_r(\RS)              & = & -\frac{1}{c}(\V \x \B_0)_r + 4 \pi \sigma
\: ,
\label{eBC2}
\end{eqnarray}
where the subscripts $r, \theta, \phi$ denote vector 
components in a spherical coordinate system,
$\V$ is the velocity of oscillation of the NS surface,
${\bf B_0}$ is the surface magnetic field inside the NS, 
$\sigma$ and $\J$ are the induced surface charge and current density.
Here (\ref{eBC2}) and (\ref{bBC2}) are used to determinate the
surface charge and current densities.
Close to the NS the current flows along the magnetic field lines. 
So, for the inner parts of the magnetosphere it can be expressed as
\be
{\bmath j} =  \alpha(r,\theta,\phi) \B
\: ,
\label{phys_current}
\ee
where $\alpha$ is a scalar function. 
This system must be completed by equation (\ref{force-free}), defining the GJ
electric field. We failed to find an analytical solution of the system in
general case. But under some physical assumptions it can be solved  analytically
for an arbitrary oscillation mode and magnetic field configuration of the NS.

Let us assume that the physical current density in the magnetosphere is low
enough -- the magnetic field to  first order in $\xiDL \equiv \xi/\RS$ can be
considered as generated only by volume currents inside the NS and by surface
currents on its surface, i.e.
\be
\frac{4 \pi}{c} {\bmath j} \ll \n \x \B^{(1)}\: ,
\label{j_ll_rot_B}
\ee
where $\B^{(1)}$ is the first order in $\xiDL$ perturbation of the magnetic field.
To order of magnitude this means:
\be
j \ll \frac{1}{r} \left(B^{(0)} \frac{\xi}{\RS}\right) \, c =
\frac{B^{(0)} \omega \xi}{c \RS}\; c\; \left(\frac{c}{\omega r}\right) \: , 
\ee
where $B^{(0)}$ is the unperturbed magnetic field strength.
The value of the Goldreich-Julian charge density near the surface of the NS is,
to order of magnitude,  $\GJ (\RS) \simeq \frac{B^{(0)} V}{c \RS}$, 
where $V$ is the oscillation velocity amplitude. Thus equation (\ref{j_ll_rot_B})
implies 
\be
j \ll \GJ(\RS) \; c\; \left(\frac{c}{\omega r}\right)\: .
\label{small_current}
\ee
We call assumption (\ref{j_ll_rot_B}) the low current density approximation.
For regions of complete charge separation, where there are charged particles 
of only one sign, the maximum current density is $\GJ \, c$.
The absolute value of the Goldreich-Julian
charge density decreases with increasing $r$ and in the near zone $r\ll
c/\omega$. Consequently condition (\ref{small_current}) for 
a charge-separated solution is satisfied automatically.
Charged particles in the near zone move along magnetic field lines. Consequently 
the current flowing through a field line tube remains the same. Because of this,
condition (\ref{small_current}) is automatically satisfied along field lines
which have crossed a region of charge separation.
For regions in the near zone, where
there are charged particles of different sign and
the magnetic field lines have not crossed a charge separated region, condition
(\ref{small_current}) {\it can} be violated.

We assume condition (\ref{small_current}) is satisfied in the whole near zone
and find the GJ electric field and the GJ charge density. For some oscillation
modes $\GJ$ 
obtained under this assumption has singularities. Because of the reasons discussed
in  Section~%
\ref{small_current_approx_section}
a regular solution of the system
(\ref{force-free}, \ref{max1}-\ref{phys_current}) should exist for any oscillation
mode and unperturbed magnetic field configuration of a NS. Hence, in cases where
{\it our} solution has singularities, the low current density approximation fails
and the physical current can not be neglected in the whole near zone. In some
regions the current density will be to order of magnitude
\be
j \simeq \GJ (\RS)\, c\, \left(\frac{c}{\omega r}\right)
\: .
\label{strong_current}
\ee
This situation will be considered in subsequent papers.
As we will show, the small current density approximation holds in regions of open
field lines. In the whole near zone it is valid for more than 50\% of the modes, 
at least for toroidal oscillation of a NS with a dipole magnetic field.

\subsection{Goldreich-Julian charge density.}

\subsubsection{Equations for Goldreich-Julian charge density}

In low current density approximation one can neglect current term in
Ampere's low (\ref{max04}).  
To the first order of dimensionless oscillation amplitude $\xiDL$ in the near
zone we have: 
\be
\n \x \B   =  0
\: .
\label{max4}
\ee
Using the properties of solenoidal vector fields  we write $\B$ in the form 
\be
\B = \n \x \n \x (P \N) + \n \x (Q \N) \: ,
\label{Bpt}
\ee
where $\N = {\bf r}/r$. Scalar functions $P$ and $Q$ can be 
expanded in spherical harmonics $Y_{lm}$ as
\begin{eqnarray}
P({\bf r}, t) & = & \sum_{lm} \tilde{P}_{lm}(r, t)\: Y_{lm}(\theta, \phi)
\label{Pexp}\\
Q({\bf r}, t) & = & \sum_{lm} \tilde{Q}_{lm}(r, t)\: Y_{lm}(\theta, \phi)\: .
\label{Qexp}
\end{eqnarray}
Substituting $\B$ in the form (\ref{Bpt}) into the equation (\ref{max4}) and
multiplying the result by $\N$ we get 
\be
\D_{\Omega} Q = 0\: ,
\label{Q0}
\ee
where $\D_{\Omega}$ is the angular part of the Laplacian,
\be
\D_{\Omega} \equiv \frac{1}{\sin \theta} 
             \pd_{\theta}( \sin \theta\: \pd_{\theta} ) +
          \frac{1}{\sin^2 \theta}\: \pd_{\phi \phi}\: .
\ee   
Substitution of the expansion (\ref{Qexp}) into the 
equation (\ref{Q0}) gives us
$\tilde{Q}_{lm} \equiv 0$ for each $l,m$. Hence, $Q \equiv 0$ 
(see~\cite{MT86}) and the magnetic field can 
be expressed in terms of only one scalar function $P$ as
\be
\B = \n \x \n \x (P \N)\: .
\label{Bp}
\ee
Substituting $\B$ from expression
(\ref{Bp}) in the Faraday`s law (\ref{max2}) we get
\be
\n \x \E = - \frac{1}{c} \n \x \n \x (\pd_t P \N) \: .
\ee
Integrating this equation:
\be
\E = - \frac{1}{c} \n \x (\pd_t P \N) - \n \Psi \: ,
\label{Egeneral}
\ee
where $\Psi$ is an arbitrary scalar function.

From the theory of partial differential equations 
\cite{Elsgolts65,Brandt47} it is known
that  for any vector field ${\bmath A}$ there exist a
vector field perpendicular to ${\bmath A}$ if and only if
\be
{\bmath A} \cdot ( \n \x {\bmath A} ) = 0\: .
\label{Pfaff}
\ee
From equation (\ref{max4}) it follows that magnetic $\B$ satisfies
equation (\ref{Pfaff}) and there always exist a vector field perpendicular to
$\B$. Let us assume that there exists an {\it electric} field $\eGJ$ 
with no component parallel to $\B$ satisfying the boundary conditions 
(\ref{eBC1}),(\ref{eBC2}).
Evidently it satisfies Maxwell equations (\ref{max1})-(\ref{max4})
and consequently has the form
\be
\eGJ = - \frac{1}{c} \n \x (\pd_t P \N) - \n \pGJ \: .
\label{E_GJ} 
\ee
The electric field $\eGJ$ satisfies equation (\ref{force-free}). 
Substituting $\eGJ$ from the expression (\ref{E_GJ}) 
in the equality (\ref{force-free}) we have an equation for $\pGJ$
\be
\frac{1}{c} \n \x (\pd_t P \N)\cdot \B  + \n \pGJ \cdot \B =0 \: .
\label{pGJ_Eqv}
\ee
Substituting $\eGJ$ from the expression (\ref{E_GJ}) into equality
(\ref{RhoGJ}) we get an expression for the GJ charge density in terms of GJ
potential $\pGJ$: 
\be
\GJ = - \frac{1}{4\pi} \D \pGJ\: .
\label{RhoGJpsi}
\ee
In other words, if one uses representation (\ref{Bp})
for a magnetic field in the near zone, 
then the Goldreich-Julian electric field
is written as a sum of two terms (\ref{E_GJ}).
The first one is the vacuum term, and the second one  ($ - \n \pGJ$) 
represents the contribution of the charged  particles in the magnetosphere 
of the NS.
The potential $\pGJ$ is a solution of the equation (\ref{pGJ_Eqv}).
In spherical coordinates, equation (\ref{pGJ_Eqv}) becomes 
(using expression [\ref{Bp}]):
\be
  \D_{\Omega} P \; \pd_r\pGJ -
  \pd_r \pd_\theta P \; \pd_\theta \pGJ -
  \frac{1}{\sin^2\theta}\: \pd_r \pd_\phi P \; \pd_\phi \pGJ + 
  \frac{1}{c \sin\theta}
      \left( \pd_r \pd_\phi P \; \pd_\theta \pd_t P -
             \pd_r \pd_\theta P \; \pd_\phi \pd_t P
      \right)  =  0\: .
\label{EquationGeneralExact}
\ee

Let us consider small oscillations of a NS,  $\xiDL \ll 1$.
We expand the function $P$ in a series in $\xiDL$
and approximate it by the sum of the first two terms:
\be
P(t, r, \theta, \phi) \approx 
      P_0(r, \theta, \phi) + \delta P(t, r, \theta, \phi)\: ,
\label{P0_plus_DeltaP}
\ee
where the function $P_0(r, \theta, \phi)$ 
is responsible for the unperturbed magnetic field and 
$\delta P(t, r, \theta, \phi)$ 
is the first order term in an expansion of $P(t, r,\theta, \phi)$ in 
$\xiDL$. 
We expand $\delta P$ in series of spherical harmonics
\be
\delta P = \sum_{l,m}\delta \tilde{p}_{lm}(t, r)\, Y_{lm}(\theta, \phi)
\: .
\label{DeltaPexp}
\ee
Subsequently we will need only the time-derivative of $\delta P$.
After substitution of the expansion (\ref{DeltaPexp}) 
in the equation (\ref{max4}) we find (see Appendix~\ref{P_and_deltaP})
\be
\pd_t\, \delta P = \sum_{l,m}\left(\frac{\RS}{r}\right)^l
                   \pd_t \delta p_{lm}(t)\, Y_{lm}(\theta, \phi)\: .
\label{dt_Delta_P}
\ee
The coefficients  $\pd_t \delta p_{lm}(t)$ 
are fixed by the boundary conditions.

\subsubsection{Boundary conditions\label{sect_boundary_conditions}}

The boundary conditions for the potential $\pGJ$ are obtained 
from the boundary conditions for electric field (\ref{eBC1}).
The tangential components of the GJ electric field on the surface of the NS
are
\be
\egj_\theta |_{r=\RS}  = 
 \left.  
   \frac{1}{c} \frac{1}{r} 
      \left( \frac{1}{r}\D_\Omega P\; v_\phi +
             \frac{1}{\sin\theta}\: \pd_r \pd_\phi P\; v_r 
      \right)
 \right|_{r=\RS}
\label{eBC1i}
\ee
\be
\egj_\phi|_{r=\RS} = 
- \left. 
    \frac{1}{c} \frac{1}{r} 
       \left( \frac{1}{r}\D_\Omega P\; v_\theta +
              \pd_r \pd_\theta P\; v_r
       \right)
   \right|_{r=\RS}
\: .
\label{eBC2i}
\ee
From the expression (\ref{E_GJ}) the tangential components 
of the GJ electric field outside the NS are
\be
\egj_\theta  = 
 - 
   \frac{1}{r} 
      \left( \frac{1}{c}\, \frac{1}{\sin\theta}\: \pd_\phi \pd_t P +
             \pd_\theta \pGJ
      \right)
\label{eBC1o}
\ee
\be
\egj_\phi  = 
   \frac{1}{r} 
        \left( \frac{1}{c} \pd_\theta \pd_t P -
               \frac{1}{\sin\theta}\: \pd_\phi \pGJ
        \right)
\: .
\label{eBC2o}
\ee
Equating these expressions on the surface of the NS and
eliminating $\pGJ$ one obtains boundary conditions for the $\theta-$ and 
$\phi-$ derivatives of $\pGJ$   
\be
\pd_\theta \pGJ|_{r=\RS} = 
- \frac{1}{c}
  \left. \left( \frac{1}{r} \D_\Omega P\; v_\phi               +
                \frac{1}{\sin\theta}\: \pd_r \pd_\phi P\; v_r      +
                \frac{1}{\sin\theta}\: \pd_\phi \pd_t P
         \right)
  \right|_{r=\RS}
\label{PsiTheta}
\ee
\be
\label{PsiPhi}
\pd_\phi \pGJ|_{r=\RS} = 
\frac{1}{c}
  \left.\left( \frac{1}{r} \sin\theta\, \D_\Omega P\; v_\theta  +
               \sin\theta\, \pd_r \pd_\theta P\; v_r      +
               \sin\theta\, \pd_\theta \pd_t P
        \right)
  \right|_{r=\RS}
\: .
\ee
From these boundary conditions we obtain the boundary condition 
to be applied to
$\delta p_{lm}(t)$. 
Differentiating equation (\ref{PsiTheta}) with respect to $\phi$
and equation (\ref{PsiPhi}) with respect to $\theta$ and equating
results we get an expression for
$\delta p_{lm}(t)$ to first order in~$\xiDL$ (see Appendix~\ref{dt_deltaPlm})
\be
\pd_t \delta p_{lm}(t) = 
\frac{1}{l(l+1)} \int_{4\pi} \, \dd \Omega\; Y^*_{lm}
\left.
  \left[ \V \cdot \n(\D_\Omega P_0) +
          \D_\Omega P_0\,(\n \cdot \V_\perp) +
          r^2 (\n_\perp(\pd_r P_0) \cdot \n_\perp)\, v_r
  \right]
\right|_{r = \RS}
\: .
\label{dt_Delta_p}
\ee
Boundary condition for $\pGJ$ can be obtained by integrating
equation (\ref{PsiTheta}) or equation (\ref{PsiPhi}).
For convenience we will use as boundary condition the result of
integrating equation (\ref{PsiTheta}) over~$\theta$. 
For a perturbation depending on time $t$ as $e^{-\I\omega t}$ 
we get to first order in $\xiDL$
\be
\pGJ|_{r=\RS}  = 
 - \frac{1}{c}
       \int 
         \dd \theta \,
         \left.
            \left(   \frac{1}{r} \D_\Omega P_0\, v_\phi          
                    + \frac{1}{\sin\theta}\: \pd_r \pd_\phi P_0\; v_r
                    + \frac{1}{\sin\theta}\: \pd_\phi \pd_t \delta P
            \right) 
         \right|_{r=\RS}
                              + e^{-\I\omega t} F(\phi)\: ,
\label{PsiBC}
\ee
where $F$ is a function of $\phi$.

For each oscillation mode the corresponding velocity field is continuously 
differentiable. From the boundary condition for 
the electric field (\ref{eBC1})
it follows that the tangential components of $\eGJ$ are finite. 
The vacuum term on the 
the right hand side in expression (\ref{E_GJ}) for the electric field $\eGJ$  
is also finite (with the natural assumption that $P$ is continuously 
differentiable with respect to $\theta$ and $\phi$ expressions 
(\ref{dt_Delta_p}) and (\ref{dt_Delta_P}) are finite for $r=\RS$).
Let us consider the azimuthal component of GJ electric field
$\egj_\phi$ near the poles%
\footnote{points with $\theta=0(\pi), r=\RS$ }.
In expression (\ref{eBC2o}) for $\egj_\phi$ outside 
the NS both $\egj_\phi$ and
the first (vacuum term) on the right side of expression (\ref{eBC2i}), 
are finite. Consequently the second term 
($\left. - \frac{\pd_\phi \pGJ}{\sin\theta}\right|_{r=\RS}$) 
is also finite, hence
$\pd_\phi \pGJ|_{\{\theta = 0,\pi;\: r=\RS\}} = 0 $.
Thus in the boundary condition (\ref{PsiBC}) one must choose $F$
such that 
$\pGJ|_{ \theta = 0,\pi;\: r=\RS } = C_{1,2}\: e^{-\I\omega t}$, 
where $C_{1,2}$ are some constants. Using the gauge freedom we  choose
\be
\pGJ|_{ \theta = 0;\: r=\RS } = 0 \: .
\label{PsiPole}
\ee

Using approximation (\ref{P0_plus_DeltaP}) for $P$  we write 
an equation for $\pGJ$ in the case of small oscillations
\be 
  \D_{\Omega} P_0 \; \pd_r\pGJ - 
  \pd_r \pd_\theta P_0 \; \pd_\theta \pGJ -
  \frac{1}{\sin^2\theta} \pd_r \pd_\phi P_0 \; \pd_\phi \pGJ +
  \frac{1}{c \sin\theta}
      \left( \pd_r \pd_\phi    P_0 \; \pd_\theta \pd_t \delta P -
             \pd_r \pd_\theta  P_0 \; \pd_\phi   \pd_t \delta P
      \right) = 0\: .
\label{EquationGeneral}
\ee
The first-order partial differential equation
(\ref{EquationGeneral}) together with the boundary conditions (\ref{PsiBC}),
(\ref{PsiPole}) and $\pd_t \delta P$ described 
by formulas (\ref{dt_Delta_P}), (\ref{dt_Delta_p}) determines $\pGJ$. 
The solution of this problem provides  $\pGJ$ to first 
order in $\xiDL$. The Goldreich-Julian charge density $\GJ$
can then be obtained from the formula (\ref{RhoGJpsi}).

\subsection{Rotating NS -- Pulsar}

At the end of this section we consider an important particular case -
a rotating neutron star. We choose $z$-axis to be parallel to the rotation axis.
In this case the partial derivative $\pd_t$ can be replaced by 
$ - \Omega\, \pd_\phi$, where $\Omega$ is the angular velocity. Thus
\be
\pd_t P = - \Omega\, \pd_\phi P \: .
\label{Pulsar_dtP}
\ee
Equation (\ref{EquationGeneralExact}) for $\pGJ$ takes the form
\be 
  \D_{\Omega} P \; \pd_r\pGJ - 
  \pd_r \pd_\theta P \; \pd_\theta \pGJ -
  \frac{1}{\sin^2\theta} \pd_r \pd_\phi P \; \pd_\phi \pGJ -
  \frac{\Omega}{c}\frac{1}{\sin\theta}
      \left( \pd_r \pd_\phi    P \; \pd_\theta \pd_\phi P -
             \pd_r \pd_\theta  P \; \pd_{\phi\phi} P
      \right)  =  0\: .
\label{PulsarEquationGeneral}
\ee
By direct substitution of
\be
\pGJ ^{\rm rot} = -\frac{\Omega}{c} \sin\theta \, \pd_\theta P
\label{Pulsar_pGJ}
\ee
it can be shown that the potential (\ref{Pulsar_pGJ})
is a solution of the equation (\ref{PulsarEquationGeneral}), 
and satisfies the boundary conditions (\ref{PsiTheta}, \ref{PsiPhi}).
We note, that equation (\ref{EquationGeneralExact}) and its particular form
(\ref{PulsarEquationGeneral}), and boundary conditions (\ref{PsiTheta},
\ref{PsiPhi})  are valid for an arbitrary oscillation amplitude, i.e. also for
rotation of the NS.
Substituting potential $\pGJ ^{\rm rot}$ into the expression for the GJ electric
field (\ref{E_GJ}) one gets
\be
\eGJ _{\rm rot} =  -\frac{1}{c} ({\bf \Omega} \x {\bmath r}) \x \B \: .
\label{Pulsar_E}
\ee
This result was obtained by  Goldreich and Julian~\shortcite{GJ69}, and 
Mestel~\shortcite{Mestel71}.

\section{Goldreich-Julian charge density and 
electromagnetic energy losses \\
of a neutron star with dipole magnetic field.
Small amplitude toroidal oscillation.
}

\subsection{General Formulas}

As an application of the developed formalism we consider the case of small
amplitude toroidal oscillations of a NS with dipole magnetic field.
Velocity field on the NS's surface for a toroidal oscillation mode $(l,m)$ is
described by \cite{Unno79}:
\begin{eqnarray}
v_r      =  0, \quad
v_\theta =  
   e^{- \I\omega t}\: W(r)\: \frac{1}{\sin\theta} \pd_\phi
                                              Y_{lm}(\theta,\phi), \quad
v_\phi   =                                         
 - e^{- \I\omega t}\: W(r)\: \pd_\theta Y_{lm}(\theta, \phi)
\: ,
\label{Vtoroidal_main}
\end{eqnarray}
where $W$ is transversal velocity amplitude.
For simplicity we assume that the mode axis is aligned with the dipole
moment ${\bmath \mu}$. We can do this without loss of generality because any
 oscillation mode with mode axis not aligned with the dipole moment 
can be represented by series of oscillation modes 
with  mode axis parallel to ${\bmath \mu}$.
In this case the unperturbed magnetic field is
\be
\B =         B_0\left(\frac{\RS}{r}\right)^3 \cos \theta\, \N +
     \frac12 B_0\left(\frac{\RS}{r}\right)^3 \sin \theta\, {\bf e_\theta}
\: ,
\label{DipB}
\ee 
where $\N$ and ${\bf e_\theta}$ are unit coordinate vectors.
This field is described  by the scalar function $P_{0}^{\rm dip}$ (see
eq. [\ref{Bp}], [\ref{P0_plus_DeltaP}])
according to the formulas (\ref{A7}), (\ref{A8}) :
\be
P_0^{\rm dip} = \frac{B_0 \RS^3}{2 r} \cos\theta
\: .
\label{DipP0}
\ee
Scalar function $\delta P^{\rm dip}$ (see eq. [\ref{Bp}]) describes first order 
in $\xiDL$ magnetic field perturbation. The time derivative of this
function for an oscillation mode $(l,m)$, according formulas 
(\ref{dt_Delta_P}), (\ref{A_delta_p_toroidal}), is
\be
\pd_t \delta P^{\rm dip}_{lm} = 
B_0 w \RS \frac{1}{l(l+1)} \left( \frac{\RS}{r}
                          \right)^l\; \pd_\phi Y_{lm}
\: ,
\label{Dip_dt_deltaP}
\ee
where $w \equiv e^{- \I\omega t} W$.
Substituting $P_0^{\rm dip}$ and $\pd_t \delta P^{\rm dip}_{lm}$ into the
general equation (\ref{EquationGeneral}) we get a partial differential equation
for GJ potential $\pGJ^{lm}$ for a NS with dipole magnetic field
oscillating with a small amplitude in a toroidal mode $(l,m)$:
\be
2 \cos\theta\; \pd_r \pGJ^{lm} + 
\frac{1}{r} \sin\theta\; \pd_\theta \pGJ^{lm} -
\frac{m^2}{l(l+1)} \frac{B_0 w}{c} \left( \frac{\RS}{r} 
                                  \right)^{l+1}\; Y_{lm}      = 0
\: ,
\label{DipEquationGeneral}
\ee
where $\pGJ^{lm}$ corresponds to one excited oscillation mode $(l,m)$, 
and for general mode $\pGJ=\sum_{lm} \pGJ^{lm}$.
The characteristics of the equation (\ref{DipEquationGeneral}) are
\begin{eqnarray}
   t & = & C_0 \nonumber \\
\phi & = & C_1 \nonumber \\
\sin\theta \left(\frac{\RS}{r}\right)^{1/2}                   & = & C_2\\
\pGJ^{lm} - 
 \frac{m^2}{l(l+1)} \frac{B_0 w \RS}{c} \left( \frac{\RS}{r}
                                        \right)^l 
\sin^{2l}\theta \; \int \frac{Y_{lm}}{(\sin\theta)^{2l+1}}\, \dd \theta & = & C_3
\nonumber 
\end{eqnarray}
The integral  of the equation (\ref{DipEquationGeneral}) is
an arbitrary function of constants $C_0, C_1, C_2, C_3$
\be
\varphi(C_0, C_1, C_2, C_3) = 0,
\ee
Expressing $\pGJ^{lm}$ we have for the general solution of 
equation (\ref{DipEquationGeneral})
\be
\pGJ^{lm} = 
    \frac{m^2}{l(l+1)} \frac{B_0 w \RS}{c} \left( \frac{\RS}{r}
                                           \right)^l 
       \sin^{2l}\theta \; \int \frac{Y_{lm}}{(\sin\theta)^{2l+1}}\, \dd \theta   +
    \Phi_{lm}\left( \sin\theta \left( \frac{\RS}{r} \right)^{1/2},
                    \phi,
                    t 
            \right)
\; ,
\label{Dip_pGJ}
\ee
where $\Phi$ is an arbitrary function. In order for 
$\pGJ^{lm}$ represented by the expression (\ref{Dip_pGJ}) to be
a GJ potential it must satisfy the boundary conditions on the surface 
of the NS (\ref{PsiBC}) and (\ref{PsiPole}), which in the case of toroidal
oscillations of a NS with dipole magnetic field  take the form
\begin{eqnarray}
\pGJ^{lm}|_{r=\RS}  & = &
- \frac{B_0 w \RS}{c}
      \int \left( \cos\theta\, \pd_\theta Y_{lm} -
                   \frac{m^2}{l(l+1)} \frac{Y_{lm}}{\sin\theta}
           \right) \, \dd \theta
                             + e^{-\I\omega t}\: F(\phi)
\: ,
\label{DipBC1}\\
\pGJ^{lm}|_{ \theta = 0, r=\RS } & = & 0
\: .
\label{DipBC2}
\end{eqnarray}
Substituting $\pGJ^{lm}$ given by the expression (\ref{Dip_pGJ}) 
into the boundary conditions (\ref{DipBC1}), (\ref{DipBC2}) 
we get the boundary condition for the function $\Phi^{lm}$:
\begin{eqnarray}
\left.\Phi^{lm} \left( \sin\theta \left( \frac{\RS}{r} \right)^{1/2},\phi,t 
               \right)\right|_{r=\RS} & = &
- \frac{B_0 w \RS}{c}
      \int \left( \cos\theta\, \pd_\theta Y_{lm} -
                   \frac{m^2}{l(l+1)} \frac{Y_{lm}}{\sin\theta}
           \right) \, \dd \theta -
\nonumber \\
&&
\phantom{+}
  \frac{B_0 w \RS}{c} \frac{m^2}{l(l+1)} 
      \sin^{2l}\theta \; \int \frac{Y_{lm}}{(\sin\theta)^{2l+1}}\, \dd \theta + 
\nonumber \\
&&
\phantom{+}
  \frac{B_0 w \RS}{c}
       \left. \left[
          \int  \left( \cos\theta\, \pd_\theta Y_{lm} -
                       \frac{m^2}{l(l+1)} \frac{Y_{lm}}{\sin\theta}
                \right) \, \dd \theta  
              \right] 
       \right|_{\theta = 0} 
\: ,
\label{DipPhiBC}
\end{eqnarray}
where the last term is added in order to satisfy the second
boundary condition (\ref{DipBC2}).
The function $\Phi^{lm}$ depends on $\theta$ in the combination 
`$\left( \frac{\RS}{r} \right)^{1/2} \sin\theta$`. To get
$\Phi^{lm}$ one has to express  the right hand side of equation 
(\ref{DipPhiBC}) in terms of `$\sin\theta$` and to replace 
`$\sin\theta$` by 
`$\left( \frac{\RS}{r} \right)^{1/2} \sin\theta$`.
Substituting the function $\Phi^{lm}$ into the expression 
(\ref{Dip_pGJ}) we get the potential $\pGJ^{lm}$ for small toroidal 
oscillations of a NS with a dipole magnetic field for the mode $(l,m)$.
Using the formula (\ref{RhoGJpsi}) we get the Goldreich-Julian charge density
for that mode. We have developed a set of programs on the computer
algebra language {\small MATHEMATICA~3.0} \cite{Wolfram96}
for calculating the analytical expressions of $\GJ^{lm}$ 
according to the  algorithm described in this section. 
These programs were tested  for some $(l,m)$ by comparing results
obtained by hand and by computer, and also by checking the
condition $\eGJ \cdot \B = 0$.

\subsection{Main Results}

\subsubsection{ Goldreich-Julian charge density}

\begin{figure}
  \epsfxsize=0.8\hsize
  \centerline{\epsfbox{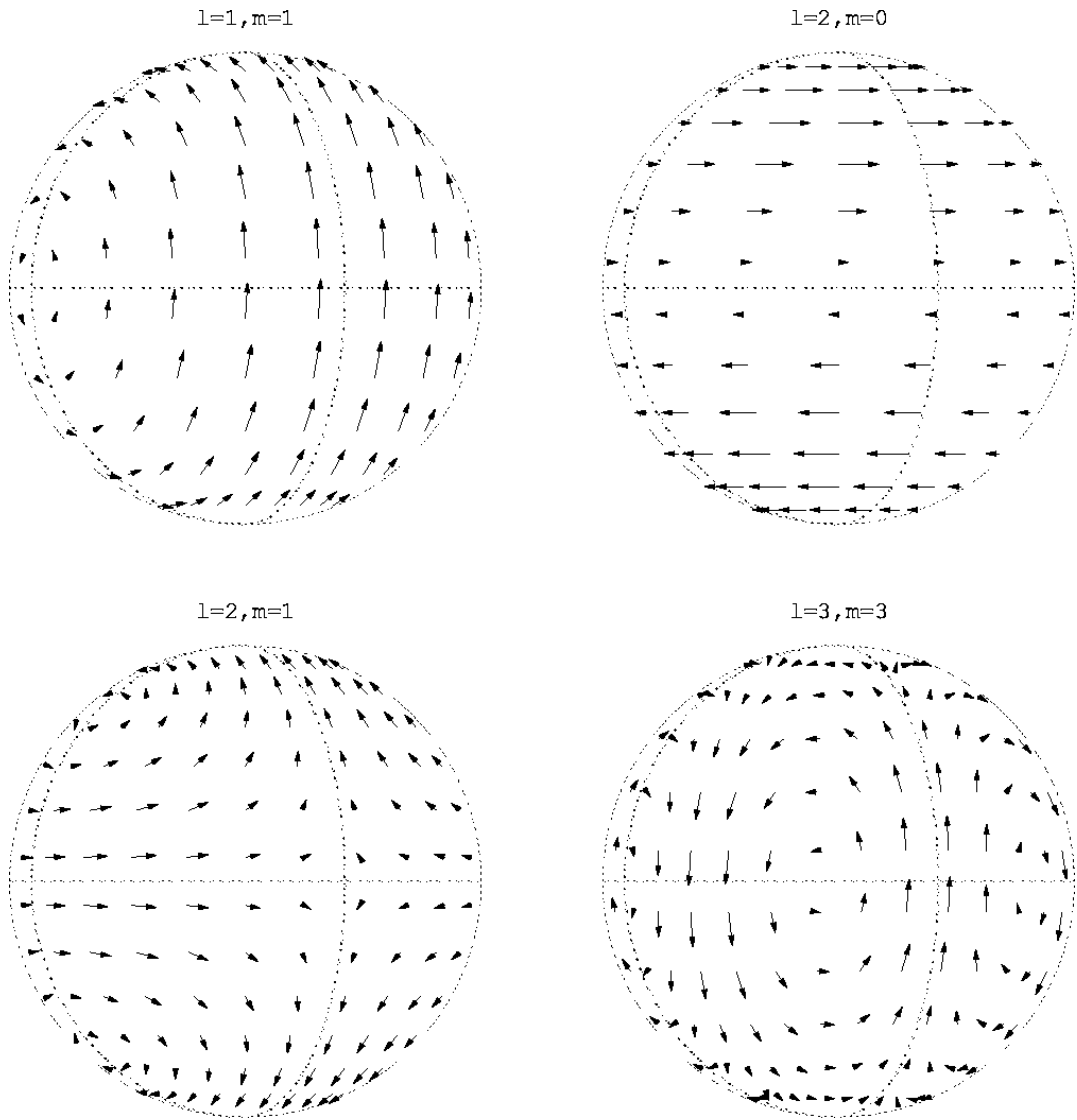}}
\caption{Velocity field on a sphere for toroidal modes 
(1,1), (2,0), (2,1) and (3,3) are shown at the time $t=2\pi\,n/\omega$, where 
$n$ is an integer, as 
projection on the meridional plane $\phi=-115^\circ$.
Circles corresponding to the longitudes $0^\circ$ (right),
-$90^\circ$ (left) and latitude $0^\circ$ are shown by dashed lines. 
}
\label{ResVel}
\end{figure}

\begin{figure}
  \epsfxsize=0.6\hsize
  \centerline{\epsfbox{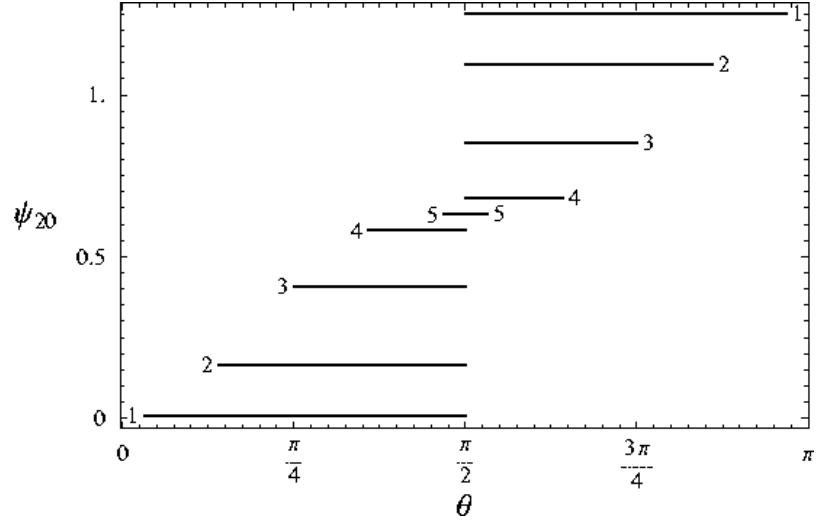}}
\caption{The potential $\pGJ^{20}$ along a dipolar magnetic field line
as a function of the polar angle $\theta$ is shown for 5 field lines
for $t=2\pi\,n/\omega$,
at which the maximal absolute values of the potential are reached.
The values of $\theta$ of the left and right ends of the lines 1-5
determine the polar angle at which corresponding
magnetic field line crosses the surface of the neutron star. 
The discontinuity is shown by the jump
of $\psi$ corresponding to the same field line in the opposite hemispheres.
}
\label{ResP20}
\end{figure}

\begin{figure}
  \epsfxsize=0.9\hsize
  \centerline{\epsfbox{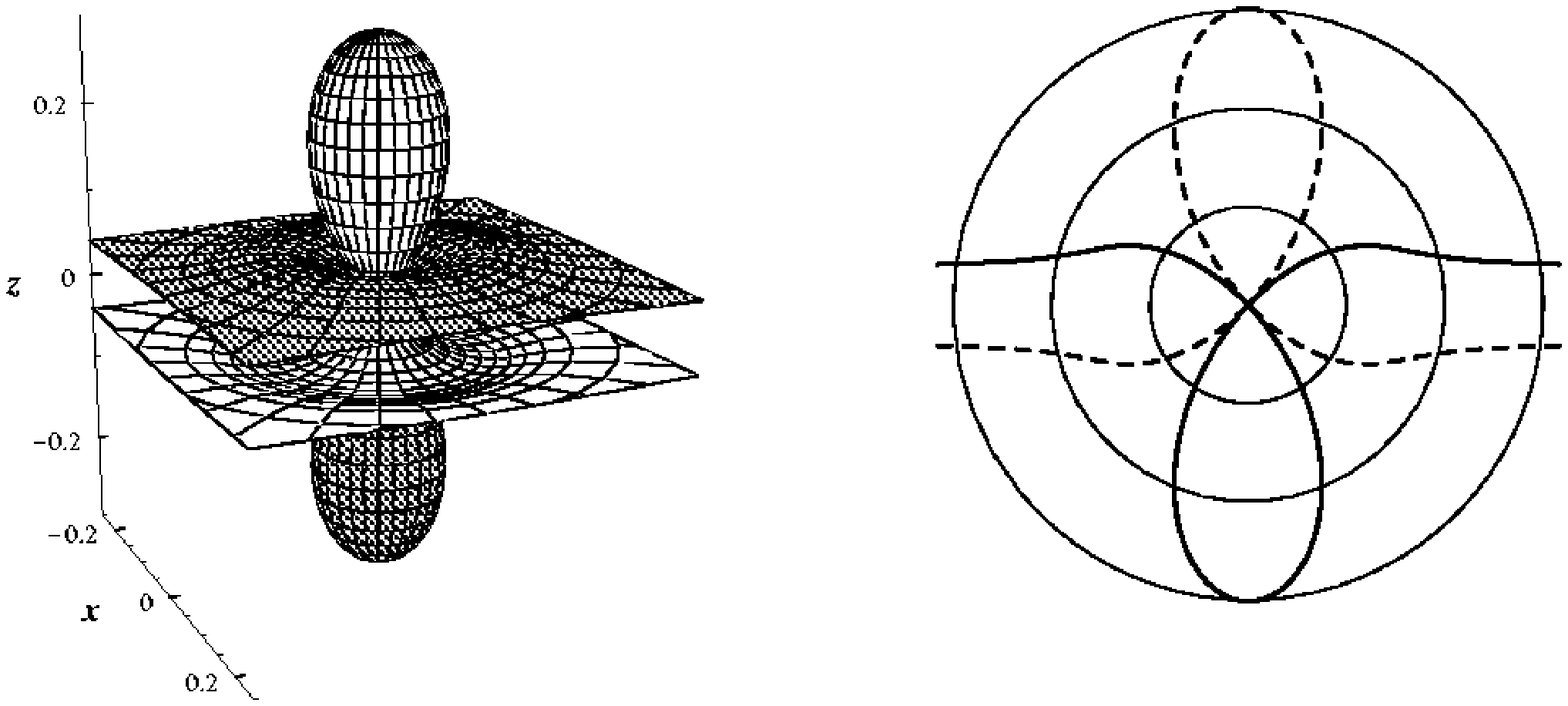}}
\caption{ 
{\it left}: 
The charge density $\GJ^{20}|_{r=\RS}$ 
is shown in a spherical coordinate system
$(\left|\GJ^{lm}|_{r=\RS}\right|, \theta, \phi)$ for $t=2\pi\,n/\omega$.
The radial coordinate represents the absolute value of the GJ charge density
as function of the polar angle $\theta$ and the azimuthal angle $\phi$.
Positive values of charge density are shown by a gray surface and 
negative ones by white. The dipole magnetic moment ${\bf\mu}$ 
is directed upwards along the $z$ axis. Azimuthal angle is counted from
the $x$-axis,  as in Fig.~\ref{ResVel}. 
Note that for $\theta=\pi/2$, $\GJ^{20}$ is infinite. 
{\it right}:
Cross-section of surfaces from the left figure by a meridional plane.
Circles correspond to the values 0.1, 0.2, and 0.3.
Positive values are shown solid, negative ones dashed.
Note that for $\theta=\pi/2$, $\GJ^{20}$ is infinite. 
}
\label{ResRGJ20}
\end{figure}

\begin{figure}
  \epsfxsize=0.6\hsize
  \centerline{\epsfbox{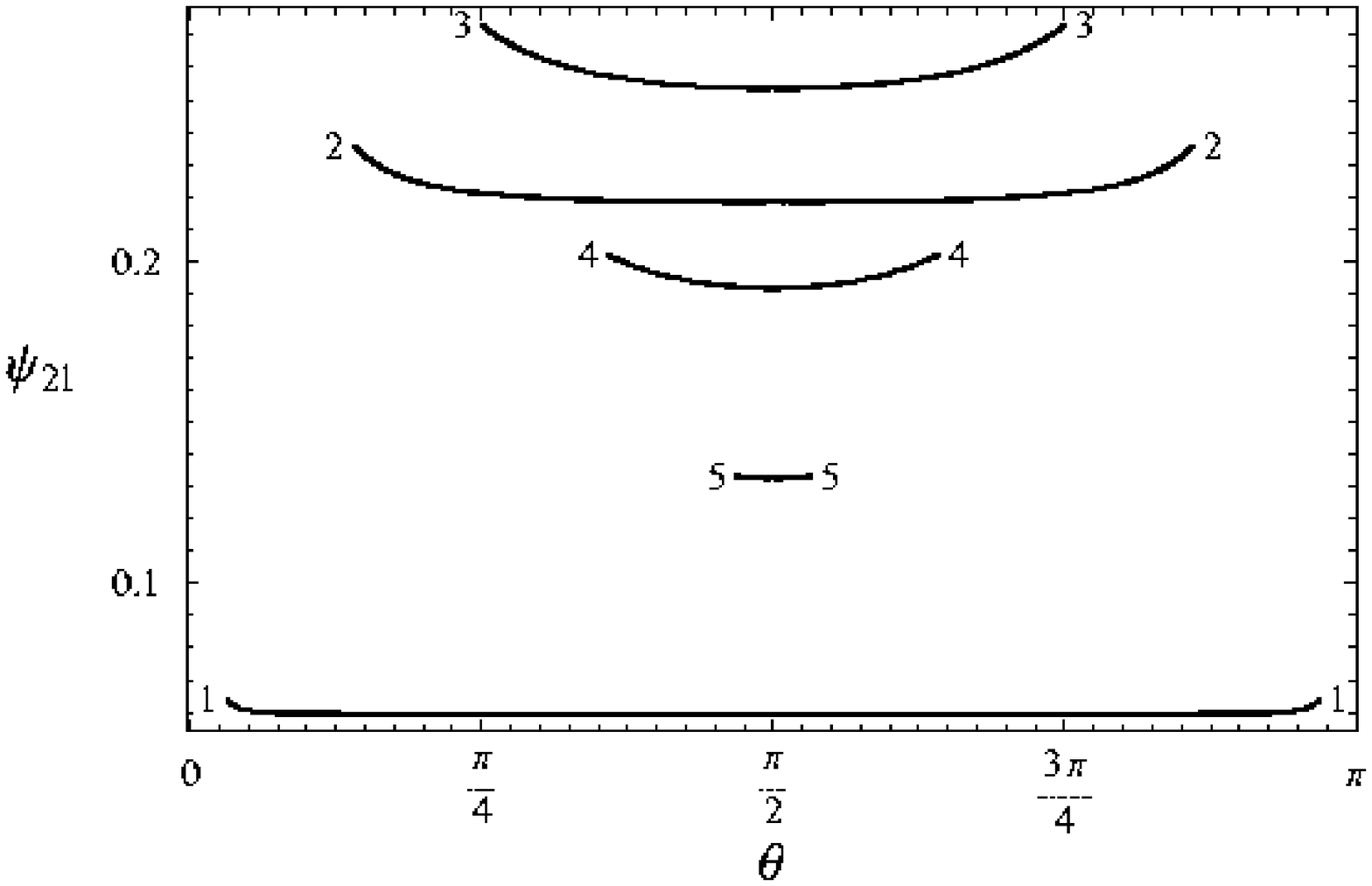}}
\caption{The potential $\pGJ^{21}$ along a dipolar magnetic field line
as a function of the polar angle $\theta$ is shown for 5 field lines
with azimuthal angle $\phi=\pi/2$, for $t=2\pi\,n/\omega$,
at which the maximal absolute values of the potential are reached.
The values of $\theta$ of the left and right ends of the lines 1-5
determine the polar angle at which corresponding
magnetic field line crosses the surface of the neutron star. 
}
\label{ResP21}
\end{figure}

\begin{figure}
  \epsfxsize=0.9\hsize
  \centerline{\epsfbox{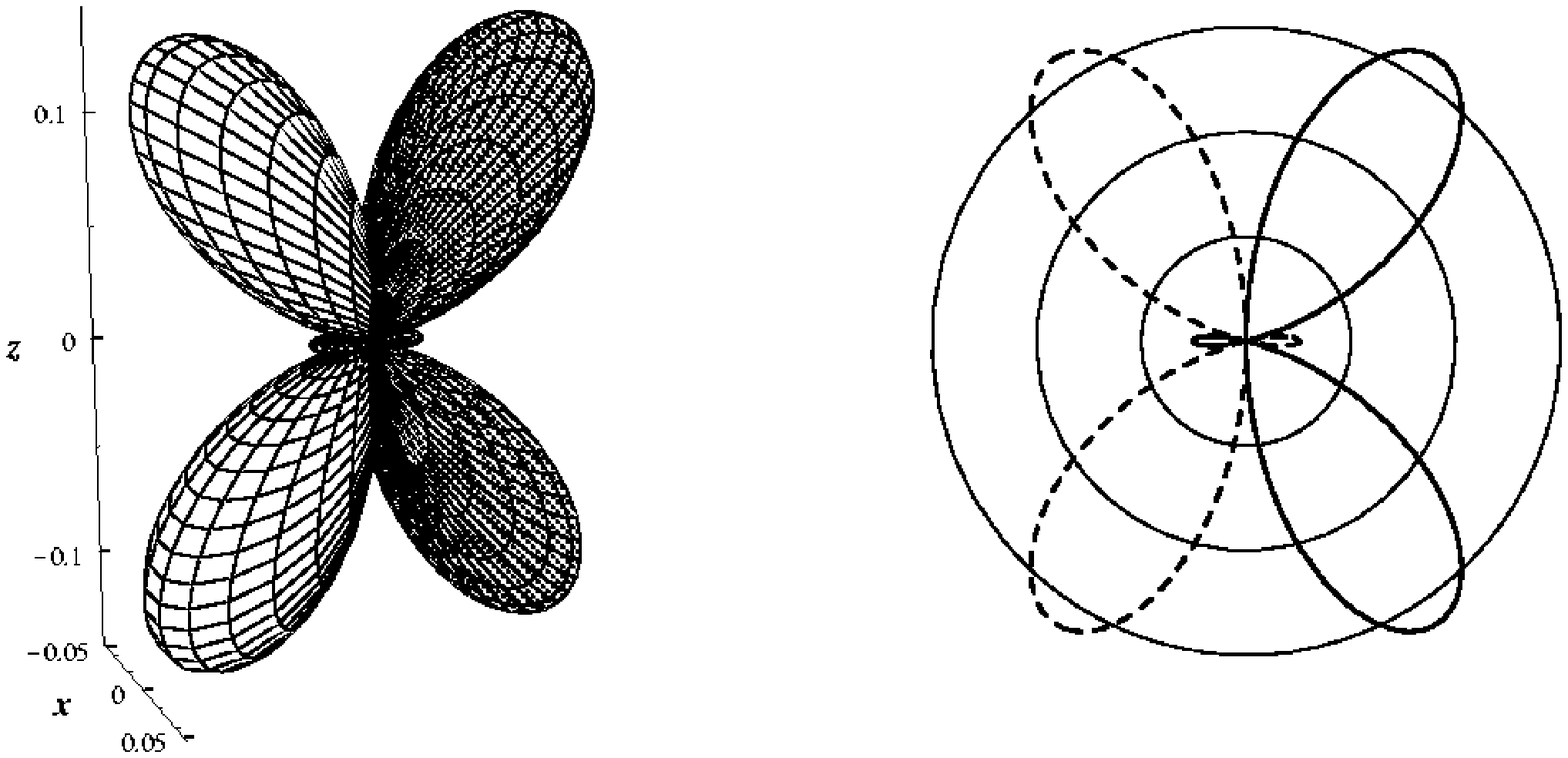}}
\caption{ 
{\it left}: 
The charge density $\GJ^{21}|_{r=\RS}$ 
is shown in a spherical coordinate system
$(\left|\GJ^{lm}|_{r=\RS}\right|, \theta, \phi)$ for $t=2\pi\,n/\omega$.
The radial coordinate represents the absolute value of the GJ charge density
as function of the polar angle $\theta$ and the azimuthal angle $\phi$.
Positive values of charge density are shown by a gray surface and 
negative ones by white. The dipole magnetic moment ${\bf\mu}$ 
is directed upwards along the $z$ axis. Azimuthal angle is counted from
the $x$-axis,  as in Fig.~\ref{ResVel}. 
{\it right}:
Cross-section of surfaces from the left figure by the plane
$\phi = \pi/2$ , where the maximal absolute values 
of the charge density $\GJ^{21}|_{r=\RS}$ are reached.
Circles correspond to the values 0.05, 0.1, and 0.15.
Positive values are shown solid, negative ones dashed.
}
\label{ResRGJ21}
\end{figure}

\begin{figure}
  \epsfxsize=0.6\hsize
  \centerline{\epsfbox{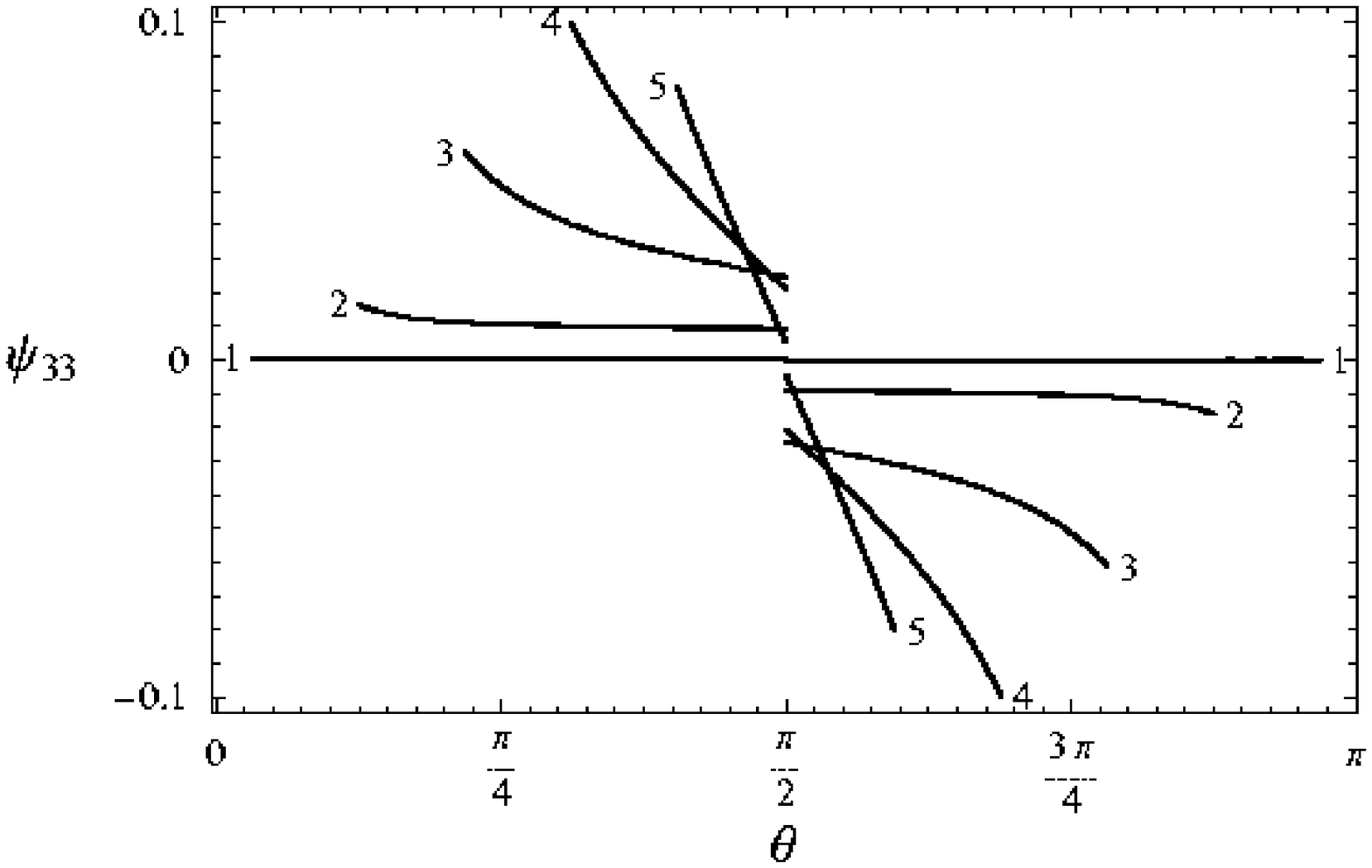}}
\caption{The potential $\pGJ^{33}$ along a dipolar magnetic field line
as a function of the polar angle $\theta$ is shown for 5 field lines
with azimuthal angle $\phi=\pi/6$, for $t=2\pi\,n/\omega$,
at which the maximal absolute values of the potential are reached.
The values of $\theta$ of the left and right ends of the lines 1-5
determine the polar angle at which corresponding
magnetic field line crosses the surface of the neutron star. 
The discontinuity is shown by the jump
of $\psi$ corresponding to the same field line in the opposite hemispheres.
}
\label{ResP33}
\end{figure}

\begin{figure}
  \epsfxsize=\hsize
  \centerline{\epsfbox{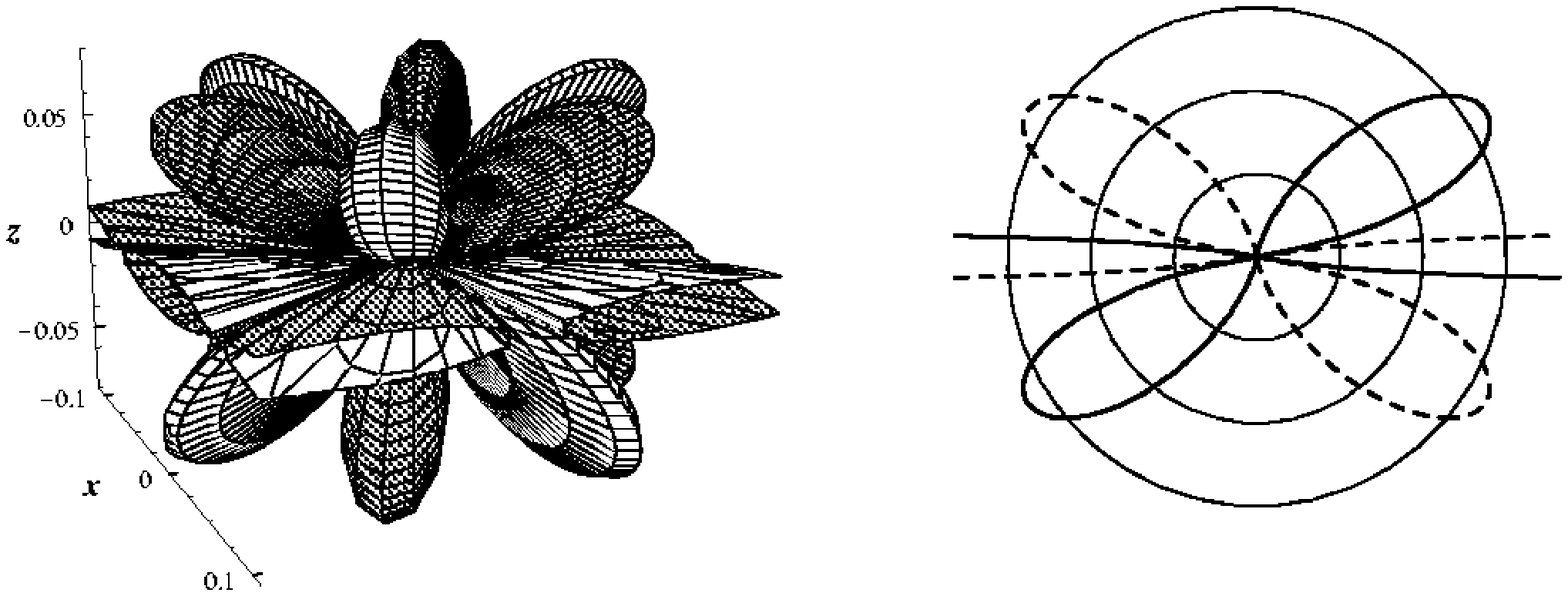}}
\caption{ 
{\it left}: 
The charge density $\GJ^{33}|_{r=\RS}$ 
is shown in a spherical coordinate system
$(\left|\GJ^{lm}|_{r=\RS}\right|, \theta, \phi)$ for $t=2\pi\,n/\omega$.
The radial coordinate represents the absolute value of the GJ charge density
as function of the polar angle $\theta$ and the azimuthal angle $\phi$.
Positive values of charge density are shown by a gray surface and 
negative ones by white. The dipole magnetic moment ${\bf\mu}$ 
is directed upwards along the $z$ axis. Azimuthal angle is counted from
the $x$-axis,  as in Fig.~\ref{ResVel}. 
Note that for $\theta=\pi/2$, $\GJ^{33}$ is infinite. 
{\it right}:
Cross-section of surfaces from the left figure by the plane
$\phi = \pi/6$ , where the maximal absolute values 
of the charge density $\GJ^{33}|_{r=\RS}$ are reached.
Circles correspond to the values 0.04, 0.08, and 0.12.
Positive values are shown solid, negative ones dashed.
Note that for $\theta=\pi/2$, $\GJ^{33}$ is infinite. 
}
\label{ResRGJ33}
\end{figure}

In the solution for the whole region the expression (\ref{DipPhiBC})
is used, where on the right hand side `$\cos\theta$` 
should be replaced by 
`$  \left(1-\left( \frac{\RS}{r} \right) \sin^2\theta \right)^{1/2}$`
for $0 \le \theta <\frac{\pi}{2}$ (first hemisphere) and by 
`$- \left(1-\left( \frac{\RS}{r} \right) \sin^2\theta \right)^{1/2}$`
for $\frac{\pi}{2} \le \theta \le \pi$ (second hemisphere). 
Because of this there are two different expressions
for $\pGJ$ for both hemispheres. 
If these expressions give different results for
$\theta=\frac{\pi}{2}$ for $r > \RS$ there is a discontinuity in
the function $\pGJ$ at the equatorial plane, and consequently
 $\GJ(r,\frac{\pi}{2},\phi)$ becomes infinite. 
As we discussed in section~%
\ref{small_current_approx_section}
this unphysical result indicate that low current density approximation for
such oscillation modes cannot be applied in the whole near zone. In this
situation it is impossible to cancel accelerating electric field in the whole
near zone without presence of a strong electric current (\ref{strong_current}) in
some magnetospheric regions. 
Examples of such oscillation modes are modes $(l,m)=(2,0),\, (3,3)$. 
The corresponding velocity fields are given in Fig.\ref{ResVel}.
The dependence of the potential $\pGJ^{lm}$
for these modes along dipole magnetic field lines is shown 
in Fig.\ref{ResP20},\ref{ResP33}. In the following we normalize
the G-J potential $\pGJ$ and charge density $\GJ$ by 
$B_0 W \RS/c$ and $B_0 W/\RS c$ respectively.
The jump of $\pGJ(r, \pi/2, \phi)$ decreases,
as the angle  at which corresponding magnetic field line intersect 
the NS`s surface is increasing. 
$\pGJ$ is a continuous function of $\theta$ on the surface of the NS,
nevertheless GJ charge density diverges on the equator
even on the surface of NS (see Fig.\ref{ResRGJ20}, \ref{ResRGJ33}).
In these figures the charge density $\GJ^{lm}|_{r=\RS}$ 
is shown in spherical coordinates  
$(\left|\GJ^{lm}|_{r=\RS}\right|, \theta, \phi)$.
Here the radial coordinate represents absolute 
value of the GJ charge density
as function of the polar angle $\theta$ and the azimuthal angle $\phi$.
Analytical expressions for the GJ charge density for the discussed modes
are given in Appendix~\ref{GJfor_some_modes}.
At the equatorial plane $\GJ^{lm}$ is infinite.
The mode $(2,1)$ is a representative of another class of oscillation modes,
in which $\pGJ^{lm}$ is a continuous function of $\theta$ 
(see Fig.\ref{ResP21}) and $\GJ^{lm}$ is finite everywhere.
The velocity field and $\GJ^{21}$ are shown in 
Figs.\ref{ResVel}, \ref{ResRGJ21}. For such modes it is  possible to
cancel longitudinal electric field 
without generating strong current along magnetic field lines.

\begin{figure}
  \epsfxsize=0.6\hsize
  \centerline{\epsfbox{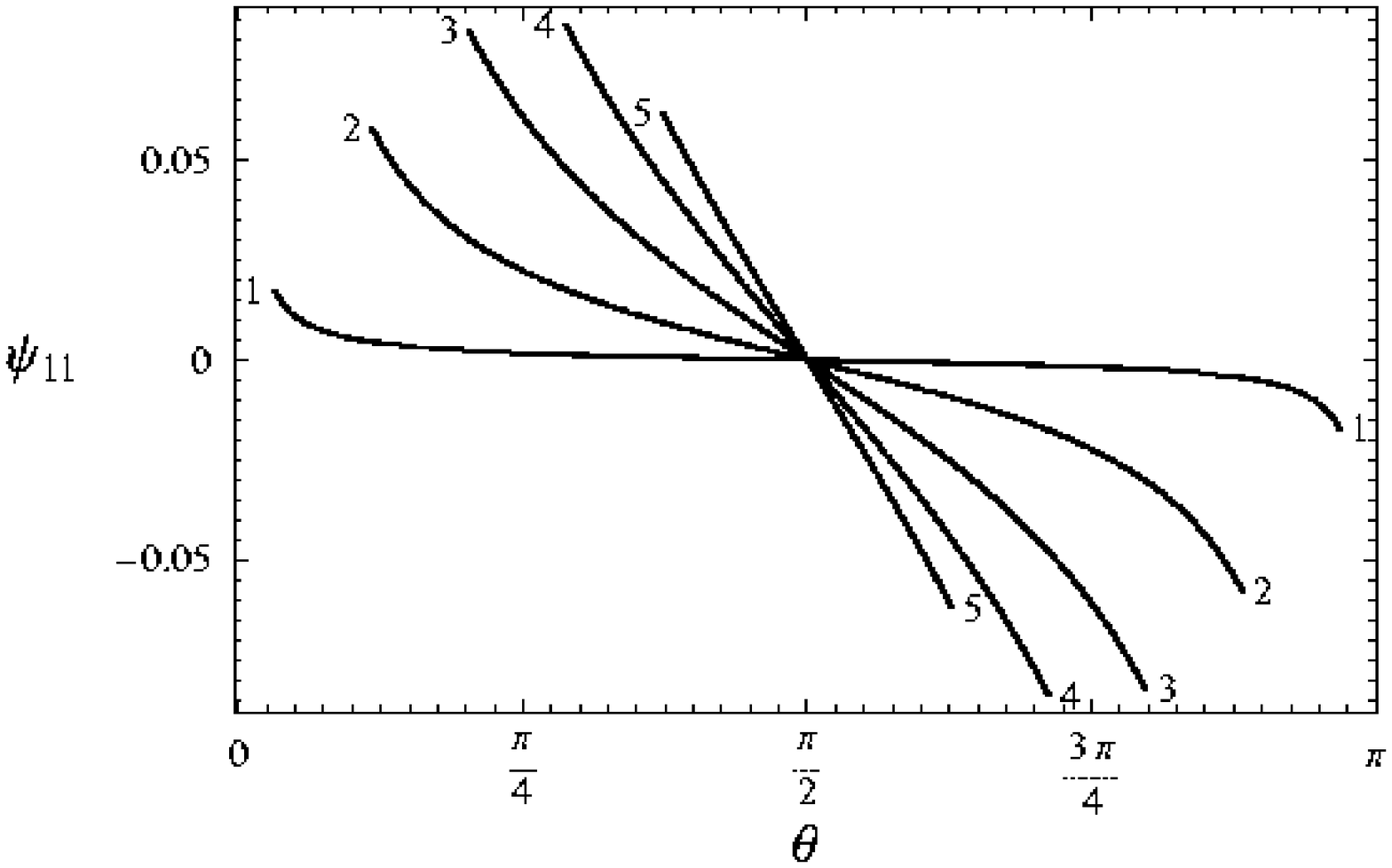}}
\caption{The potential $\pGJ^{11}$ along a dipolar magnetic field line
as a function of the polar angle $\theta$ is shown for 5 field lines
with azimuthal angle $\phi=\pi/2$, for $t=2\pi\,n/\omega$,
at which the maximal absolute values of the potential are reached.
The values of $\theta$ of the left and right ends of the lines 1-5
determine the polar angle at which corresponding
magnetic field line crosses the surface of the neutron star. 
}
\label{ResP11}
\end{figure}

\begin{figure}
  \epsfxsize=0.9\hsize
  \centerline{\epsfbox{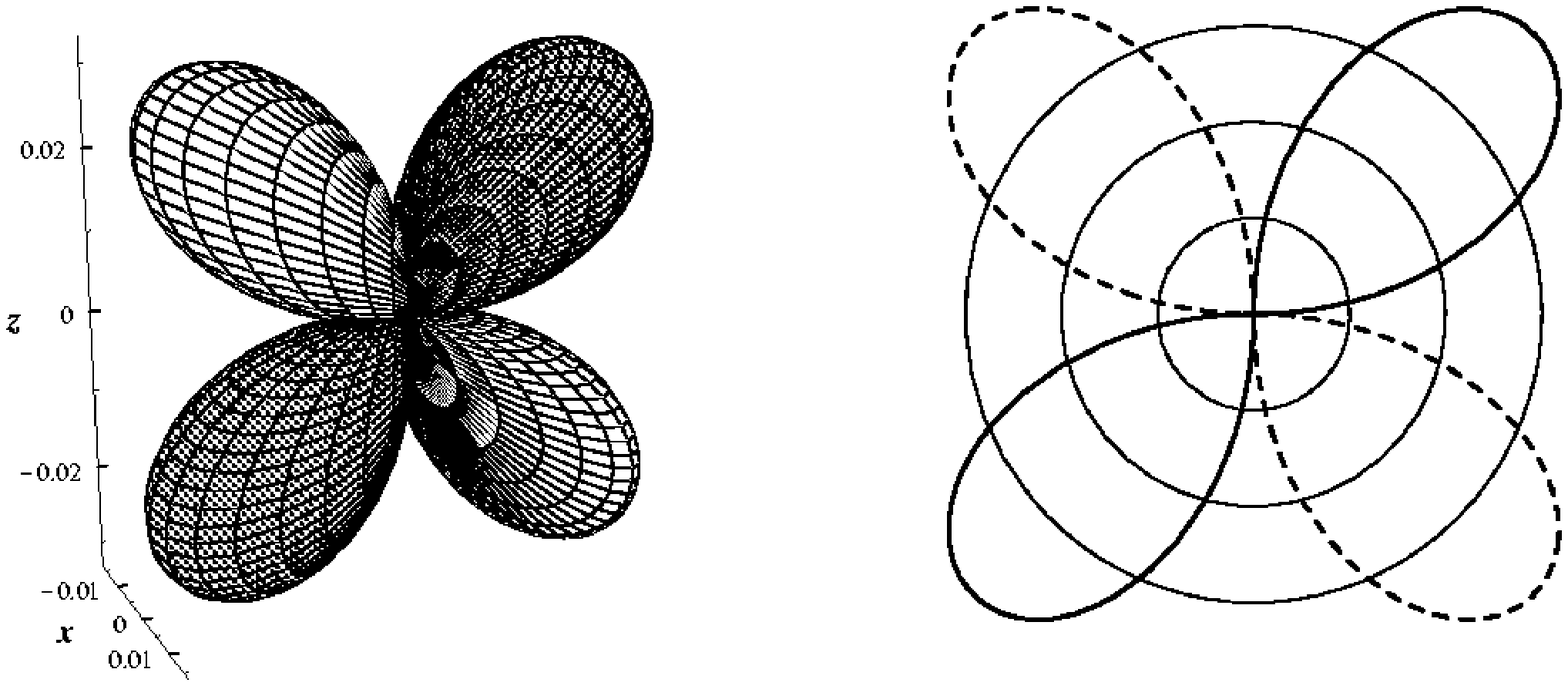}}
\caption{ 
{\it left}: 
The charge density $\GJ^{11}|_{r=\RS}$ 
is shown in a spherical coordinate system
$(\left|\GJ^{lm}|_{r=\RS}\right|, \theta, \phi)$ for $t=2\pi\,n/\omega$.
The radial coordinate represents the absolute value of the GJ charge density
as function of the polar angle $\theta$ and the azimuthal angle $\phi$.
Positive values of charge density are shown by a gray surface and 
negative ones by white. The dipole magnetic moment ${\bf\mu}$ 
is directed upwards along the $z$ axis. Azimuthal angle is counted from
the $x$-axis,  as in Fig.~\ref{ResVel}. 
{\it right}:
Cross-section of surfaces from the left figure by the plane
$\phi = \pi/2$ , where the maximal absolute values 
of the charge density $\GJ^{11}|_{r=\RS}$ are reached.
Circles correspond to the values 0.01, 0.02, and 0.03.
Positive values are shown solid, negative ones dashed.
}
\label{ResRGJ11}
\end{figure}

Consider now the difference between two classes of
toroidal oscillation modes for a NS with a dipole magnetic field.
The boundary condition for $\pGJ^{lm}$ is proportional to the vector product
of $\V$ and $\B$. 
The axis $z$ is directed along the dipole magnetic momentum.
If the velocity field is symmetric relative to the equatorial plane
($l-m$ is an odd number), then the boundary condition for $\pGJ^{lm}$
is also symmetric relatively to the equatorial plane. 
Hence, after expression of all
trigonometric functions in (\ref{DipPhiBC}) through `$\cos\theta$` and 
`$\sin\theta$` , `$\cos\theta$` appears in the right hand side of
(\ref{DipPhiBC}) only in even powers. Consequently the function $\Phi^{lm}$
will be the same in both hemispheres and $\GJ^{lm}$ will be finite.
If velocity field is antisymmetric relative to the equatorial plane
($l-m$ is an even number or zero), then `$\cos\theta$` is contained 
in the right side of (\ref{DipPhiBC}) in odd powers, and the
function $\Phi^{lm}$ may have  discontinuity in the plane
$\theta = \frac{\pi}{2}$, and $\GJ^{lm}$ for such modes
becomes infinite at the equatorial plane. 
An example of oscillation mode
with velocity field antisymmetric relative to the equatorial plane 
and for which $\pGJ^{lm}$ is continuous function of $\theta$
is the mode $(1,1)$. Graphics for the potential $\pGJ^{11}$, 
Goldreich-Julian charge density $\GJ^{11}$
are shown in Figs. \ref{ResP11}, \ref{ResRGJ11}. 
So, for a dipole magnetic field the longitudinal electric field generated by
the oscillation modes with odd $(l-m)$ can be canceled by putting
of charged particles in the magnetosphere.
Among the modes with even or zero $(l-m)$ there are modes, for which 
longitudinal electric field cannot be canceled by putting
of charged particles without presence of strong electric 
current along some magnetic field lines.

For an oscillation mode with odd $(l-m)$ there is a regular solution for $\GJ$
and $\eGJ$ in the low current density approximation. This solution will be stable
because of Lenz's low: increasing of the current will lead to a generation of a
magnetic field inducing an electric field, which will prevent the current to
grow. In other words, configuration with the smallest possible current will take place.
For oscillation modes with even or zero $(l-m)$, where no regular solution for GJ 
charge density and electric field exists, the low current density approximation
cannot be used in the whole near zone. Because some of oscillation modes of
this class (at least one) possess regular solution in low current density
approximation, the 
total amount of modes, for which this approximation may be used in the whole
near zone exceeds 50\%.

The Goldreich-Julian charge density gives only a characteristic 
charge density in the magnetosphere. The particle density
can be obtained by solving full system of MHD equations, but this has not
been done even for an aligned rotator with a dipole magnetic field 
\cite{Michel91}. The case of nonradial
oscillation is much more complicated, because of the absence of
stationarity and axial symmetry. 
About particle number density in the region of closed magnetic field
lines we can say, that for oscillation modes with
odd $(l-m)$ there can be charge separated regions in the near zone.
For such modes the foot points of magnetic field lines
have the same sign of the potential and there are regions where GJ charge
density does not change the sign along the whole magnetic filed line.  Hence,
charged particles of only one sign 
can be pulled from the NS surface into these regions. 
Evidently for modes with even or zero $(l-m)$
there can not be charge-separated regions in the near zone, because GJ charge
density change the sign along any magnetic field line.

\subsubsection{Energy Losses\label{ResEnergyLosses}}

In previous works the calculation of electromagnetic energy losses
of an oscillating neutron star includes only radiation of 
electromagnetic waves in the vacuum. 
If there is plasma in the magnetosphere, then the energy 
will be lost by a transformation of the oscillation energy into
kinetic energy of an outflowing plasma, as it was proposed for 
pulsar energy losses by Goldreich and Julian \shortcite{GJ69}. 
They obtained the energy losses of a rotating aligned NS through outflow 
of the charged particles from a region of open field lines, 
and found it to be equal to the loss in vacuum through
radiation of electromagnetic waves by a perpendicular rotator.
In this paper we consider electric and magnetic fields in
the near zone, hence we can not explicitly show the existence of
plasma outflow as in the aligned rotator by Goldreich and
Julian, but a qualitatively the picture is the following. 
Electric current arising from the motion of the
charged particles in the magnetosphere generate a magnetic field.
In the region where the value of this field becomes larger than 
the unperturbed NS magnetic field the field lines become open, 
on the Alfvenic surface.
Plasma escaping from the region of open field lines 
produces an electromagnetically
driven stellar wind. The wind causes a net electric current which closes
at infinity. This current flows beneath the stellar surface between 
positive and negative emission regions. Because it must cross magnetic 
field lines there, it exerts a braking torque on the
oscillating NS and reduces the oscillational energy.
This picture is similar to the one proposed by 
Ruderman \& Sutherland \shortcite{Ruderman75} for pulsars, 
but now we must get the boundary of
the open field line region self-consistently. In our case we can not
approximate the closed field line region  by a ``zone of corotation'', rather
we should self-consistently determine the last closed field line 
as the last field line lying inside the Alfven surface.

Consider region near the pole.  Let us denote as $\theta_0$ 
a polar angle by which the
last closed field line intersects the surface of the NS.
Similar to pulsars above a polar cap an acceleration zone 
and a zone of $e^+ -e^-$ pair generation above it will be builded.
Electron-positron pairs cancel accelerating electric field, i.e. motion of charged
particle above the accelerating zone will be not influenced by 
electric filed generated by stellar oscillations.
In the region of open field lines, where particles escape from the
neutron star and form relativistic wind, the average time of crossing 
of the acceleration zone by the particle is much less then the
oscillation period (see Discussion), 
so particles practically do not return back to the star.
Hence, averaged over the oscillation period, the energy loss through 
the outflow of plasma from open field line region is 
\be
\epsilon^{lm}_{\rm pl} \simeq \frac{1}{\tau}\int_0^\tau\, \dd t\,
                \int_0^{2\pi} \int_0^{\theta_0}
                     \left|
                        j^{\, lm}(\RS,\theta,\phi) \: A^{lm}(\theta,\phi)
                     \right|\; 
                  \RS^2 \sin\theta\, \dd\theta \dd\phi
\: ,
\label{epsilon}
\ee
where $A(\theta,\phi)$ is the work done by the electric field to move a unit
charge to the point with coordinate~$(\RS,\theta,\phi)$:
\be
A(\theta,\phi) \simeq 
\int_0^\theta  \egj_\theta(\RS, \theta^\prime, \phi)\: 
               \RS\, \dd\theta^\prime
\: .
\label{work}
\ee
Because of reasons discussed at the end of section~%
\ref{small_current_approx_section}
charge density in the inner parts of the magnetosphere must be
approximately equal to the Goldreich-Julian charge density $\GJ$.
In the region of plasma outflow charged particles are streaming along
magnetic field lines in the same direction with ultrarelativistic velocities.
Charge density of outflowing plasma is equal to the Goldreich-Julian
charge density. Hence, outflowing ultrarelativistic electron-positron plasma
builds a net current density
\be
j \simeq \GJ\; c\: .
\label{current}
\ee
The current density can differ from the values given by
expression~(\ref{current}) due to difference in averaged velocities of
electron and positron components of the plasma (discussion on 
this topic see in Lyubarskii \shortcite{Lyubarskii92}), but on the order of magnitude
expression~(\ref{current}) gives good estimation for the current density 
\cite{Ruderman75}. Hence, for the open field line region
condition~(\ref{small_current}) is satisfied and in
expression~(\ref{current}) we can use $\GJ$ obtained by solving of
equation~(\ref{EquationGeneral}) for any oscillation mode.

The last closed field line is the line for which the  
kinetic energy density of  outflowing plasma on equator 
(at the point $(R_{\rm a}, \pi/2, \phi)$)
becomes equal to the corresponding energy density of the NS magnetic field:
\be
\frac{\epsilon^{lm}_{\rm pl}(\theta_0)}{4 \pi R_{\rm a}^2 c} \simeq
\left. \frac{B^2}{8\pi}\right|_{\theta=\frac{\pi}{2}, r=R_{\rm a}}
\: .
\label{epsilon_equal_B}
\ee
Expressing the right hand side of this equation in terms of
angle $\theta_0$, we get two equations (\ref{epsilon}),
(\ref{epsilon_equal_B}) for a self-consistent determination
of $\theta_0$ and $\epsilon^{lm}_{\rm pl}$. 

\begin{figure}
  \epsfxsize=0.7\hsize
  \centerline{\epsfbox{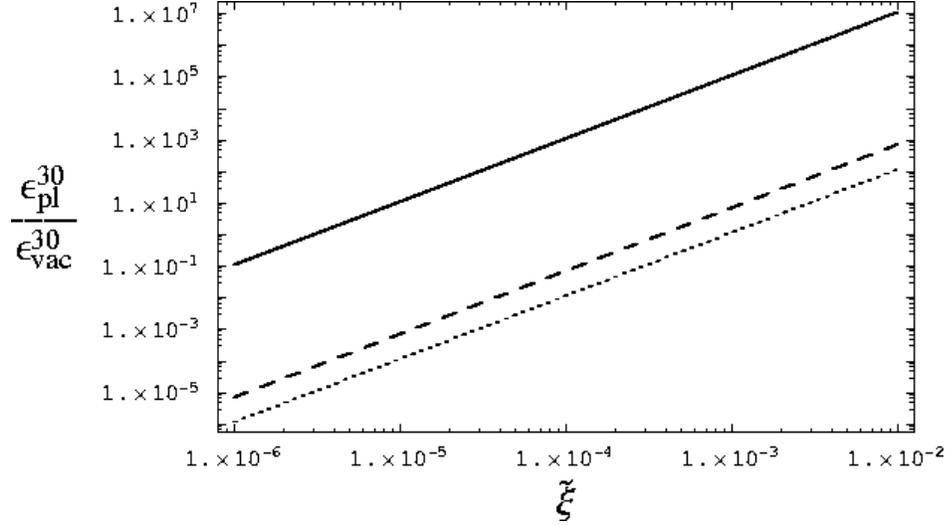}}
\caption{ Ratio of the energy loss through outflow of plasma
from the region of open field lines to the loss in vacuum
 is shown as a function 
of the dimensionless displacement amplitude $\xiDL$ for the mode
(3,0).
The solid line corresponds the oscillation period $T = 19 ms$
(mode $_3t_0$ in notations of McDermott et al. (1988)), the
dashed line -- $T = 1.7 ms$ (mode $_3t_1$),
the dotted line -- $T = 1.08 ms$ (mode $_3t_2$); 
see note on the page~\pageref{Mode_30}. 
}
\label{ResEEvac30}
\end{figure}

We have solved these equations 
for oscillation modes (1,1), (2,0), and (3,0).
Oscillation periods%
\footnote{According to McDermott et al. \shortcite{McD88} the dependence of
the eigenfrequency on $l$ is very weak, at least for small $l$, so we 
assumed the same oscillation periods for modes (2,0) and (3,0)
}\label{Mode_30} 
were taken from McDermott et al. \shortcite{McD88}. 
The ratio of energy losses
to the vacuum ones as a function of the amplitude of 
dimensionless transversal displacement $\xiDL$ for
modes (1,1), (2,0), and (3,0) are given in Table~\ref{ResTableRatios}. 
One can see that energy losses
through outflowing plasma are much less than in vacuum for small 
values of $\xiDL$. On the other hand,
for oscillation modes $l\ge 2$, for displacement amplitudes
$\xiDL$ larger than a value $\xicrit$, energy losses exceed the vacuum losses
(see Fig. \ref{ResEEvac30}).
For some oscillation modes for which $\theta_0 \ll 1$, we can linearize
equations (\ref{epsilon}), (\ref{epsilon_equal_B}) in $\theta_0$ and
get an analytical estimation of $\theta_0$.
For toroidal oscillation mode $(l,m)$ the velocity amplitude $V$
near the pole is of the order
\be
V \sim  W\, \left\{ \begin{array}{ll}
                        \theta^{m-1}, & m \neq 0  \\
                        \theta      , & m   =  0
                      \end{array}
              \right.
\label{v_phi_pg}
\ee
The electric field is of the order $\egj \sim B\, V/c$. 
From formula (\ref{work}) we have
\be
A \sim \frac{B W \RS}{c}\:
         \left\{ \begin{array}{ll}
                   \theta^m, & m \neq 0 \\
                   \theta^2, & m   =  0
                 \end{array}
         \right.
\label{work_pg}
\ee
The Goldreich-Julian charge density is
\be
\GJ \sim  \DSt\frac{B W }{\RS c}\: \theta^m
\label{rgj_pg}
\ee
Substituting expressions (\ref{work_pg}) and (\ref{rgj_pg}) into
equations (\ref{current}), (\ref{work}) and (\ref{epsilon})
we get the energy loss through the
plasma outflow along open field lines: 
\be
\epsilon_{\rm pl} \sim   B_0^2 \RS^2 c \left( \frac{W}{c} 
                                          \right)^2 
\:  
 \left\{ \begin{array}{ll}
           \theta_0^{2m+2},& m \neq 0\\
           \theta_0^4,     & m   =  0
         \end{array}
 \right.
\ee
The angle $\theta_0$ is determined from the equation 
(\ref{epsilon_equal_B}). After substitution of $\epsilon_{\rm pl}$ into
equation (\ref{epsilon_equal_B}) we get
\be
\theta_0 \sim \left\{ \begin{array}{ll}
                       \DSt\left( \frac{W }{c}
                           \right)^{\frac{1}{3-m}}, & m \neq 0  \\
\\
                       \DSt\left( \frac{W }{c}
                           \right)^{\frac{1}{2}},   & m   =  0
                      \end{array}
              \right.
\label{theta_pg}
\ee
We see that the angle $\theta_0$ is small only for modes with $m<3$.
This is because of the small value of $V$ in the polar region for
modes with $m \ge 3$. 
For such  modes a linear analysis is not possible and 
the case of large $\theta_0$ needs additional investigation,
which will be given elsewhere.
The energy loss through plasma outflow is
\be
\epsilon_{\rm pl} \sim B_0^2\, \RS^2\, c\: 
        \left\{ \begin{array}{ll}
                     \DSt \xiDL^{\frac{8}{3-m}}\:
                         \left( \frac{\RS \omega }{c}
                         \right)^{\frac{8}{3-m}}, & m = 1,2  \\
\\
                     \DSt \xiDL^4\:
                         \left( \frac{\RS \omega }{c}
                         \right)^4,               & m =  0 
                   \end{array}
         \right.
\label{ResPplasm}
\ee 
The vacuum energy loss for the oscillation mode $(l,m)$, according to
McDermott et al. \shortcite{McD84} is
\be
\epsilon_{\rm vac}\sim 
B_0^2\, \RS^2\, c\: \xiDL^2 
            \left\{\begin{array}{ll}
                   \DSt\left(\frac{\RS\omega}{c}\right)^8,      & l=1, m=0 \\
\\
                   \DSt\left(\frac{\RS\omega}{c}\right)^{2l+2}, & l=1, m=1;\,
                                                            l\ge 2, m=0\dots l\\
                   \end{array}
            \right.
\label{ResPvac}
\ee
From these formulas we get the oscillation amplitude $\xiDL$
for which the plasma energy loss is larger than the vacuum loss
\be
\xiDL >
\xicrit \sim 
  \left\{ \begin{array}{ll}
            \DSt\left(\frac{\RS\omega}{c}\right)^2,       & l=1, m=0 \\
\\
            \DSt\left(\frac{\RS\omega}{c}\right)^{l-1},   & l\ge 2, m=0;\,
                                                            l\ge 1, m=1 \\
\\
            \DSt\left(\frac{\RS\omega}{c}\right)^{l/3-1}, & l\ge 2, m=2
          \end{array}
  \right.
\label{xicrit}
\ee
From formula (\ref{xicrit}) it follows that
for modes $(l=1, m=0)$, $(l \ge 2, m=0,1)$ and $(l\ge 4, m=2)$ nonvacuum 
energy loss exceeds the vacuum loss even for small displacement amplitudes 
$\xiDL$, because for these modes $\xicrit \ll 1$.

\begin{table}
\centerline{
\begin{tabular}{|c|c|c|}
\hline
Mode  & $\epsilon_{\rm pl}/\epsilon_{\rm vac}$ & $R^{lm}_{\rm a}/\lambda$ \\ 
\hline
$(1,1)$ & $ 0.02\: \xiDL^2 $                  & $ 0.9/\xiDL $ \\
\hline
$(2,0)$ & $ 77\: \xiDL^2 \tau^2 $              & $ 0.2/\xiDL $ \\ 
\hline
$(3,0)$ & $ 9 \cdot 10^5\: \xiDL^2 \tau^4 $    & $ 0.09/\xiDL $ \\ 
\hline
\end{tabular}
}
\caption{Leading terms in expansion over $\xiDL$ of ratios of electromagnetic
energy loss through plasma outflow to the vacuum one~%
$\epsilon_{\rm pl}/\epsilon_{\rm vac}$ 
and distance 
in the equatorial plane of the last closed field line to the
wave length 
$R^{lm}_{\rm a}/\lambda$ for toroidal oscillation modes $(1,1)$, $(2,0)$ 
and $(3,0)$. $\tau$ is period of oscillations in milliseconds.}
\label{ResTableRatios}
\end{table}

\begin{figure}
  \epsfxsize=0.6\hsize
  \centerline{\epsfbox{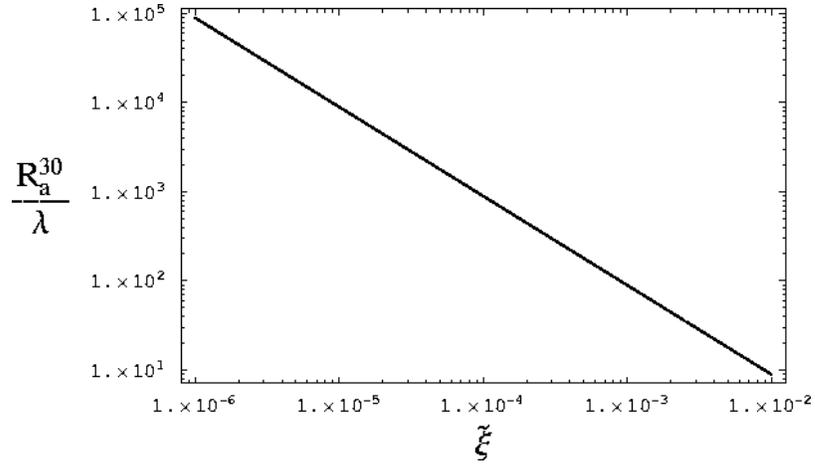}}
\caption{ Ratio of the equatorial radius of the last closed field line
to the wavelength $\lambda$ is shown as a function 
of the dimensionless displacement amplitude $\xiDL$ for the mode
(3,0). This ratio does not depend on the oscillation frequency $\omega$.
}
\label{ResRRw30}
\end{figure}

\begin{figure}
  \centerline{\epsfbox{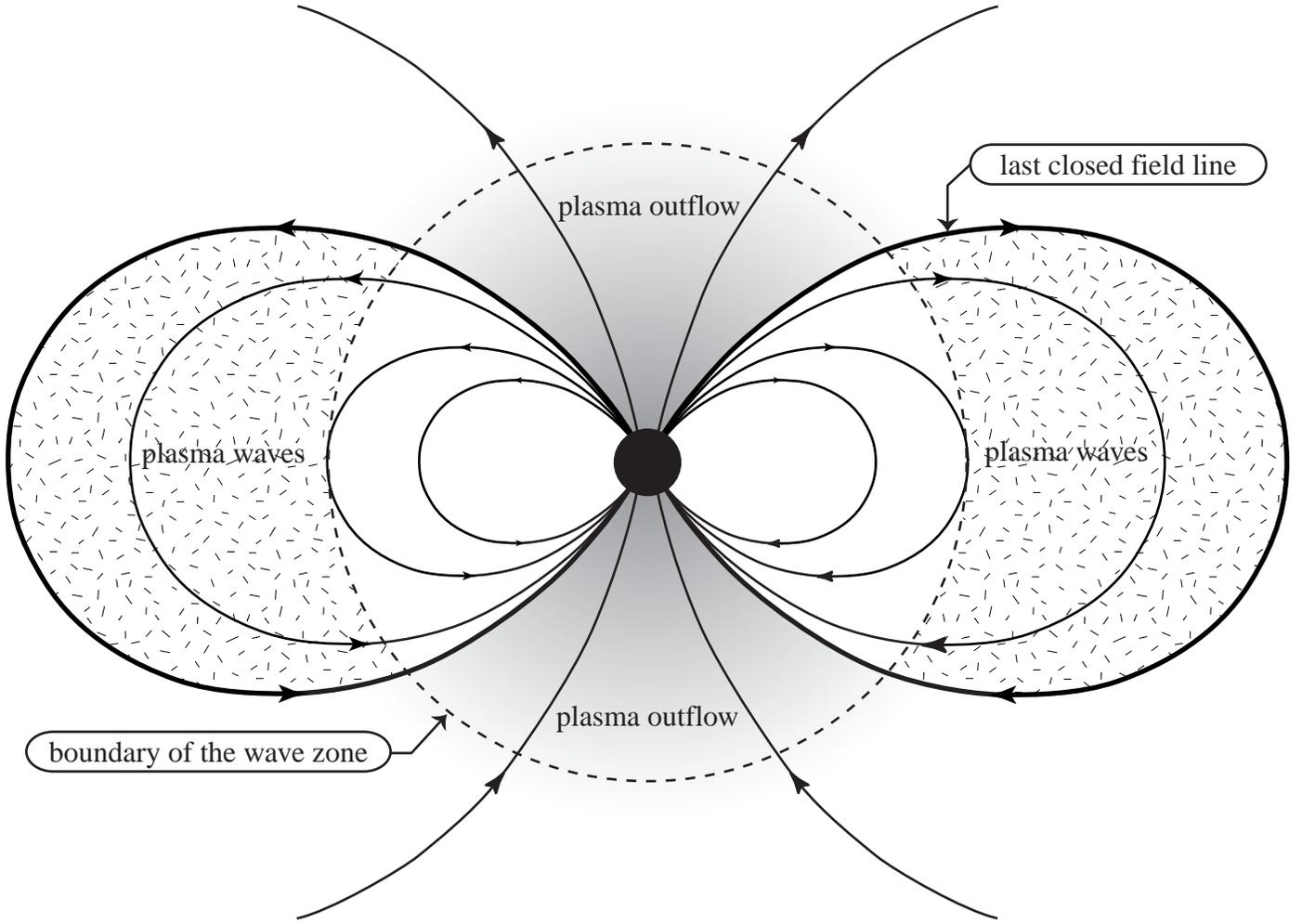}}
\caption{ Structure of the magnetosphere of an oscillating 
neutron star (arbitrary scaled). See explanations in text.
}
\label{ResNSpicture}
\end{figure}

Electromagnetic waves radiated by the oscillating NS are screened 
by the plasma. For small $\xiDL$ the nonvacuum energy loss according
to~(\ref{xicrit}) is smaller than 
the vacuum loss because at the GJ density of particles
the accelerated plasma cannot escape from
the magnetosphere. The ratio of the equatorial radius for the
last closed field line, to the wave length $\lambda$ are given in
Table~\ref{ResTableRatios}. 
For mode $(3,0)$ this dependence is shown in Fig.~\ref{ResRRw30}. 
We see that the last closed field line
for small values of displacement amplitude
lies deep inside the wave zone. In the region
limited by the last closed field lines and a sphere of radius
$\lambda$ (formal boundary of the wave zone) in this situation confined 
plasma waves are excited such as Alfven or magnetosonic wave by
the oscillation of the NS, see Fig.~\ref{ResNSpicture}. 
This will lead to accumulation of energy
in this region in the form of plasma wave energy. 
When the density of this energy becomes larger 
than the magnetic energy, field lines becomes open.
These processes will lead to a decrease of the equatorial radius 
of the last closed field lines, and hence, a increase of the energy loss.

A second process which could lead  to larger
energy losses than obtained by the solution of the equation 
(\ref{epsilon_equal_B}), is rotation of the neutron star. 
In this case the equatorial radius of the last closed 
field line is the radius of the light cylinder 
$R_{\small L} =\Omega/c$, and surface area for plasma outflow 
may be considerably increased. 
It is not clear, however, how the interaction between 
rotation an oscillation will influence the total energy losses.

\section{Discussion}

Under the assumption of low current density 
we have developed a formalism describing the inner parts of the magnetosphere of
an oscillating neutron star in the same way as in the case of rotation.
This assumption is valid in the region of open field lines for any oscillation
mode. For some oscillation modes the assumption about low current density
everywhere in the inner magnetosphere leads to unphysical results -- the
Goldreich-Julian charge density becomes infinite in some points. For such
oscillation modes the low current density approximation cannot be used in the whole 
region of closed magnetic field lines near the NS. The current density for these
modes along some closed magnetic field lines will be of the order of magnitude 
$j\simeq \frac{c}{\omega r}\, \GJ\: c$. 
For a rotating neutron star our formalism yields the same values for
the Goldreich-Julian charge density and electric field as obtained in works of
Goldreich and Julian~\shortcite{GJ69} and Mestel~\shortcite{Mestel71}. 

We applied the general formalism to the case of toroidal oscillations 
of a neutron star with dipole magnetic field. We calculated the GJ charge density
and showed, that for the case considered the assumption about
low current density in the whole inner magnetosphere is valid for more than
half of all modes.
We calculated the electromagnetic energy losses of a NS with dipole magnetic field
for some toroidal oscillation modes and showed, that the energy losses of an
oscillating neutron star are strongly affected by the plasma present in its
magnetosphere.
Electromagnetic energy losses of an oscillating NS due to plasma outflow from
the magnetosphere have in general another dependence on oscillation frequency
$\omega$ and the dimensionless displacement amplitude $\xiDL$ than the energy
losses due  
to radiation of electromagnetic waves in vacuum.
This affects all previous calculations of the electromagnetic damping rate of NS
oscillation. Our calculations give a lower limit on 
electromagnetic energy losses of an oscillating NS because of the reasons
discussed at the end of section~\ref{ResEnergyLosses}.
The energy of a star oscillation
depends on the dimensionless displacement amplitude as $\xiDL^2$. Energy losses through
plasma outflow in contrast to energy losses through radiation 
of electromagnetic waves in vacuum
have another dependence on the displacement amplitude. 
Consequently the electromagnetic energy losses of an oscillating NS 
in general depends on the oscillation amplitude $\xiDL$.
To estimate the oscillation damping time for the considered oscillation modes one
can use values given in McDermott et al. \shortcite{McD88}, Table 6,
multiplying by functions from Table~\ref{ResTableRatios} of this paper
for a given displacement amplitude.
Here we have restricted ourselves to a simple case of toroidal
oscillations. Starquakes should excite a whole spectrum of oscillations,
including $p$- and $g$- modes. The magnetosphere structure produced 
by these modes will be considered in a future paper.
 
With the proposed formalism it is possible to apply theoretical 
models developed for pulsars to oscillating neutron stars. 
For example, one can investigate the acceleration mechanism 
proposed by Scharlemann, Arons \& Fawley \shortcite{Arons79} or
by Ruderman \& Sutherland \shortcite{Ruderman75} for oscillating neutron star.
The  inertial frame dragging mechanism
proposed by Muslimov \& Tsygan \shortcite{MT92} 
does not work here, because we consider a nonrotating star.
Characteristic time of pair cascade formation is estimated as
$h_{\rm PPF}/c$, where $h_{\rm PPF}$ is the hight of pair formation front
above the surface.
In ``Ruderman-Sutherland'' model minimum value of $h_{\rm PPF}$ is 
the same as for a NS rotating with the angular frequency
$\Omega_{\rm eff} = \omega \xiDL$, and for usually assumed parameters of
NS oscillation is of order $10^4-10^5$ cm.
In ``Arons-Scharlemann'' like model the hight of pair formation front
can be smaller then in a pulsar rotating with the angular frequency
$\Omega_{\rm eff}$ (several km), because in general for nonradial oscillation
ratio $\GJ/B$ along magnetic field lines increase faster than
in magnetosphere of a rotating NS, what leads to larger accelerating 
electric fields.
Because the characteristic time of pair
cascade formation is much shorter than the period of oscillation, one
can consider the polar cap acceleration zone in any given moment of time
as stationary. Problem of return current region for
the case of oscillating NS as in the case of pulsar remains open.
Similar to pulsars return current can flow along 
the last closed magnetic field line.

\bigskip

The authors wish to thank A. I. Tsygan for helpful discussions.
This work has been supported by Russian Foundation of Basic Research
(RFFI) under grants 96-02-16553 and 99-02-18180.

\clearpage

\clearpage

\newpage
\centerline{\bf APPENDIX }

\appendix
\section{Calculation of $P$ and $\delta P$}
\label{P_and_deltaP}

Substituting $\B$ from
formula (\ref{Bp}) using expansion (\ref{P0_plus_DeltaP})
into the equation (\ref{max4}) we have
\be
\n \x \n \x \n \x (P_0 +\delta P)\N = 0 \: .
\label{A1}
\ee
Because $P$ does not depend on $t$, equation (\ref{A1}) is equivalent to 
the following two equations:
\begin{eqnarray}
\n \x \n \x \n \x (P_0 \N) = 0 \: \phantom{.}
\label{A2}\\
\n \x \n \x \n \x (\delta P \N) = 0 \: .
\label{A3}
\end{eqnarray}
In spherical coordinate equation (\ref{A3}) is
\be
\n \x \n \x \n \x (\delta P \N) =  
  - {\bf e}_\theta \frac{1}{r \sin\theta}
       \left( \frac{1}{r^2}\: \pd_\phi\, \D_\Omega \delta P +
              \pd_{rr} \pd_\phi \delta P
       \right) + 
    {\bf e}_\phi \frac{1}{r}
       \left( \frac{1}{r^2}\: \pd_\theta\, \D_\Omega \delta P +
              \pd_{rr} \pd_\theta \delta P
       \right) = 0 \: . 
\label{A4}
\ee
Substituting expansion of $\delta P$ (\ref{DeltaPexp}) we get
\be
- \frac{1}{r^2} l(l+1) \delta \tilde{p}_{lm}(r,t) +
            \pd_{rr} \delta \tilde{p}_{lm}(r,t) = 0 \: .
\label{A5}
\ee
The solution of equation (\ref{A5}) which vanishes at infinity  is
\be
\delta \tilde{p}_{lm}(r,t) = \left(\frac{\RS}{r}\right)^l \delta p_{lm}(t) 
\: .
\label{A6}
\ee
The time-dependent coefficients $p_{lm}(t)$ have to be determined
from the boundary conditions. Substituting expression (A6) 
into expansion (\ref{DeltaPexp}) and taking the derivative 
with respect to $t$ we get  expression (\ref{dt_Delta_P}).

Similarly, for the coefficients $P_{0\,lm}$ in the expansion of the function
$P_0$ we have
\be
P_{0\,lm}(r) = p_{lm} \left(\frac{\RS}{r}\right)^l
\label{A7}
\ee
Substituting the expansion of the function $P_0$ in spherical 
harmonics into the boundary condition (\ref{bBC1}), multiplying
this expression by $Y^*_{lm}$, and integrating it over solid 
angle $4 \pi$, we have 
(compare with formula (16) in Muslimov \& Tsygan \shortcite{MT86})
\be
p_{lm} =    \frac{\RS ^2}{l(l+1)}
            \int_{4 \pi} B_{0\, r} Y^*_{lm} \, \dd\Omega ,
\label{A8}
\ee
where $Y^*_{lm}$ is inverse spherical harmonic 
$\int Y_{lm} Y^*_{l^\prime m^\prime}\, \dd\Omega 
= \delta _{ll^\prime}\delta _{mm^\prime} 
$
and $\dd\Omega = \sin\theta \dd\theta\, \dd\phi$ is the solid angle

\section{Calculation of \lowercase{$\delta p_{lm}(t)$}}
\label{dt_deltaPlm}

Differentiating equation (\ref{PsiTheta}) with respect to $\phi$
and equation (\ref{PsiPhi}) with respect to $\theta$, 
equating the results and multiplying by $1/\sin\theta$ we have
\begin{eqnarray}
\left.
  \left[
    \frac{1}{\sin\theta}\pd_\theta(\sin\theta\, \pd_\theta(\pd_t P)) +
    \frac{1}{\sin^2\theta}\pd_{\phi \phi} (\pd_t P) 
  \right]
\right|_{r=\RS} & = &
-
\left\{\,
  \frac{1}{r} \D_\Omega P 
      \left[\frac{1}{\sin\theta} \pd_\theta (\sin\theta\: v_\theta ) +
             \frac{1}{\sin\theta} \pd_\phi\ v_\phi 
     \right] 
\right.
+
\nonumber\\
&&
\phantom{- \{ .}
\left.
\left.
     \left[
       \frac{1}{\sin\theta} \pd_\theta(\sin\theta\: \pd_r \pd_\theta P) +
       \frac{1}{\sin^2\theta} \pd_r \pd_{\phi \phi} P
     \right] \; v_r
+
\right.
\right.
\nonumber\\
&&
\phantom{- \{ .}
\left.
\left.
  \left[ \frac{1}{r}            \pd_\theta \D_\Omega P  \; v_\theta +    
         \frac{1}{r \sin\theta} \pd_\phi   \D_\Omega P  \; v_\phi 
  \right]
+
\right.
\right.
\nonumber\\
&&
\phantom{- \{ .}
\left.
\left.
  \left[ \pd_r \pd_\theta P\; \pd_\theta v_r +
         \frac{1}{\sin^2\theta} \pd_r \pd_\phi P\; \pd_\phi v_r
  \right]
\,
\right\}
\right|_{r=\RS}
\label{B1}
\end{eqnarray}
Simplifying equation (\ref{B1}) by writing
$P$ as a sum of the two terms (\ref{P0_plus_DeltaP}), we get an expression 
for time-dependent part $\delta P$ to first order in~$\xiDL$
\begin{eqnarray}
\left. \pd_t \D_\Omega \delta P \right|_{r=\RS} & = &
- \left[\, \left( v_r      \; \pd_r \D_\Omega P_0  +
                   v_\theta \; \frac{1}{r} \pd_\theta \D_\Omega P_0  +   
                   v_\phi   \; \frac{1}{r\sin\theta} 
                                            \pd_\phi  \D_\Omega P_0
           \right)
+  \D_\Omega P_0 
      \left(\frac{1}{r \sin\theta} \pd_\theta (\sin\theta\: v_\theta ) +
            \frac{1}{r \sin\theta} \pd_\phi\ v_\phi 
      \right) 
  \right.
+
\nonumber \\
&&
\phantom{- [ .}
\left.
\left.
   \left( \pd_r \pd_\theta P_0\; \pd_\theta v_r +
          \frac{1}{\sin^2\theta} \pd_r \pd_\phi P_0\; \pd_\phi v_r
   \right)
\,
\right]
\right|_{r=\RS}  
\end{eqnarray}
or in vector notation
\be
\left. \pd_t \D_\Omega \delta P \right|_{r = \RS} = 
-
\left.
  \left\{ \V \cdot \n(\D_\Omega P_0) +
          \D_\Omega P_0(\n \cdot \V_\perp) +
          r^2 (\n_\perp(\pd_r P_0) \cdot \n_\perp)\, v_r
  \right\}
\right|_{r = \RS},
\label{B3}
\ee
where $\V_\perp$ is the tangential part of the velocity,
$\V_\perp = {\bf e}_\theta \: v_\theta + {\bf e}_\phi\: v_\phi$,
and
$\n_\perp$ is the angular part of the gradient: 
$\n_\perp = {\bf e}_\theta \: 1/r\:\pd_\theta + 
                  {\bf e}_\phi   \: 1/(r \sin\theta)\:\pd_\phi
$.
Substituting the expansion of $\delta P$ in spherical harmonics 
(\ref{dt_Delta_P}), multiplying the result by $Y^*_{lm}$ and
integrating it over the solid angle $4 \pi$, we get
an expression for the coefficients of $\pd_t \delta p(t)$  (\ref{dt_Delta_p})
in the expansion of $\delta P$ in spherical harmonics:
\begin{eqnarray*}
\pd_t \delta p_{lm}(t) = 
\frac{1}{l(l+1)} \int_{4 \pi} \, \dd\Omega\; Y^*_{lm}
\left.
  \left[ \V \cdot \n( \D_\Omega P_0 ) +
          \D_\Omega P_0 (\n \cdot \V_\perp) +
          r^2 (\n_\perp(\pd_r P_0) \cdot \n_\perp)\, v_r
  \right]
\right|_{r = \RS}
\: .
\end{eqnarray*}

Next we give explicit expressions for coefficients $\delta p_{lm}(t)$ 
for toroidal and spheroidal oscillations modes 
(see Unno et al. \shortcite{Unno79}).
Spheroidal separation of variables gives the following expressions for
components of the oscillation velocity in spherical coordinates
\begin{eqnarray}
v_r      =  e^{- \I\omega t}\: U(r)\: Y_{lm}(\theta, \phi), \quad
v_\theta = 
   e^{- \I\omega t}\: V(r)\: \pd_\theta Y_{lm}(\theta, \phi), \quad
v_\phi   =
   e^{- \I\omega t}\: V(r)\: \frac{1}{\sin\theta} \pd_\phi 
                                              Y_{lm}(\theta, \phi)
\: ,
\label{Vspheroidal}
\end{eqnarray}
where $U$ and $V$ are radial and transversal velocity amplitude 
respectively. Substituting the velocity components
(\ref{Vspheroidal}) into formula (\ref{dt_Delta_P}), 
we get the coefficients $\pd_t \delta p_{l'm'}(t)$ for spheroidal oscillations
\begin{eqnarray}
\pd_t \delta p_{l'm'}(t) & = &
  \frac{ e^{- \I\omega t} }{l'(l'+1)}
  \int_{4\pi}\, \dd\Omega\; Y^*_{l'm'}\; 
  \left[ \,
     U 
       \left(                     Y_{lm} \; \pd_r \D_\Omega P_0  + 
                       \pd_\theta Y_{lm} \; \pd_\theta \pd_r P_0 +
  \frac{1}{\sin^2\theta }\pd_\phi Y_{lm} \; \pd_\phi \pd_r P_0 
       \right) +
  \right.
\nonumber\\
&&
\phantom{\frac{1}{l'(l'+1)}
  \int_{4\pi}\, \dd\Omega\; Y^*_{l'm'} 
[ .
}
\left.
\left.
    \frac{V}{r} 
                \left(    -l(l+1)\; Y_{lm} \; \D_\Omega P_0            + 
                       \pd_\theta Y_{lm} \; \pd_\theta \D_\Omega P_0 +
  \frac{1}{\sin^2\theta }\pd_\phi Y_{lm} \; \pd_\phi \D_\Omega P_0 
                \right)
\,
\right]
\right|_{r = \RS}
\end{eqnarray}
The oscillation velocity components for toroidal oscillations are
\begin{eqnarray}
v_r      =  0, \quad
v_\theta =  
   e^{- \I\omega t}\: W(r)\: \frac{1}{\sin\theta} \pd_\phi
                                              Y_{lm}(\theta,\phi), \quad
v_\phi   =                                         
 - e^{- \I\omega t}\: W(r)\: \pd_\theta Y_{lm}(\theta, \phi)
\: ,
\label{Vtoroidal}
\end{eqnarray}
where $W$ is transversal velocity amplitude.
With the use of formula (\ref{Vtoroidal}), the coefficients 
in the expansion of $\pd_t \delta P$ for toroidal oscillations are
\begin{eqnarray}
\pd_t \delta p_{l'm'}(t) =
  \frac{ e^{-\I\omega t} }{l'(l'+1)}
     \int_{4\pi}\, \dd\Omega\; Y^*_{l'm'}\;
 \left.
  \left[
   W 
     \frac{1}{r \sin\theta}
        \left( \pd_\phi Y_{lm} \; \pd_\theta \D_\Omega P_0 -
               \pd_\theta Y_{lm} \; \pd_\phi \D_\Omega P_0                 
        \right)
  \right] 
\right|_{r = \RS}
\: .
\label{A_delta_p_toroidal}
\end{eqnarray}

\section{Expressions for Goldreich-Julian charge density\\
         for modes (1,1), (2,0), (2,1), (3,3) }
\label{GJfor_some_modes}

\noindent 
$\GJ^{11}$ for $0 \le \theta \le \pi$
\be
\GJ^{11} = e^{-i \omega t}
\frac{3\,{\sqrt{\frac{3}{2}}}\,B\,W\,\cos (\theta )\,\sin (\theta )\,
    \sin (\phi )\,{r_*}^2}{8\,c\,{\pi }^{\frac{3}{2}}\,r^3}
\ee

\bigskip

\noindent
$\GJ^{20}$ for $0 \le \theta < \frac{\pi}{2}$ 
\be
\GJ^{20} = - e^{-i \omega t}
\frac{3\,{\sqrt{5}}\,B\,W\,{{r_*}}^2\,
    \left( 16\,r - 3\,{r_*} + 15\,\cos (4\,\theta )\,{r_*} - 
      12\,\cos (2\,\theta )\,\left( -4\,{r} + 
         {{{r}}_*} \right)  \right) }{256\,c\,
    {\pi }^{\frac{3}{2}}\,r^4\,
    {\sqrt{1 - \frac{{\sin (\theta )}^2\,{r_*}}{r}}}}
\ee
$\GJ^{20}$ for $\frac{\pi}{2} < \theta \le \pi$
\be
\GJ^{20} = e^{-i \omega t}
\frac{3\,{\sqrt{5}}\,B\,W\,{{r_*}}^2\,
    \left( 16\,r - 3\,{r_*} + 15\,\cos (4\,\theta )\,{r_*} - 
      12\,\cos (2\,\theta )\,\left( -4\,{r} + 
         {r_*} \right)  \right) }{256\,c\,
    {\pi }^{\frac{3}{2}}\,r^4\,
    {\sqrt{1 - \frac{{\sin (\theta )}^2\,{r_*}}{r}}}}
\ee

\bigskip

\noindent
$\GJ^{21}$ for $0 \le \theta \le \pi$
\be
\GJ^{21} = e^{-i \omega t}
\frac{{\sqrt{\frac{5}{6}}}\,B\,W\,\sin (\theta )\,\sin (\phi )\,
    {r_*}^{\frac{3}{2}}\,
    \left( 21\,r + 19\,{r_*} + 
      45\,\cos (2\,\theta )\,{r_*} \right) }{32\,c\,
    {\pi }^{\frac{3}{2}}\,r^{\frac{7}{2}}}
\ee

\bigskip

\noindent
$\GJ^{33}$ for $0 \le \theta < \frac{\pi}{2}$
\begin{eqnarray}
\GJ^{33} & = & e^{-i \omega t}
\frac{3\,{\sqrt{35}}\,B\,W\,{\sin (\theta )}^3\,\sin (3\,\phi )\,
    {r_*}^{\frac{5}{2}}}%
{256\,c\,{\pi }^{\frac{3}{2}}\,r^{\frac{11}{2}}\,
    {\sqrt{1 - \frac{{\sin (\theta )}^2\,{r_*}}{r}}}}
\,
    \left( \vphantom{\sqrt{\cos(2\theta)}}
      15\,r^2 + 2\,r\,{r_*} + 
      6\,\cos (4\,\theta )\,{r_*}^2 - 
\right.\\
&&
\qquad\qquad 
\left. 
      6\,\cos (2\,\theta )\,{r_*}\,
       \left( -5\,r + {r_*} \right)  - 
      6\,{\sqrt{2}}\,\cos (3\,\theta )\,
       {r_*}^{\frac{3}{2}}\,
       {\sqrt{2\,r - {r_*} + \cos (2\,\theta )\,{r_*}}} \
\right)
\end{eqnarray}
$\GJ^{33}$ for $\frac{\pi}{2} < \theta \le \pi$
\begin{eqnarray}
\GJ^{33} & = & - e^{-i \omega t}
\frac{3\,{\sqrt{35}}\,B\,W\,{\sin (\theta )}^3\,\sin (3\,\phi )\,
    {r_*}^{\frac{5}{2}}}%
{256\,c\,{\pi }^{\frac{3}{2}}\,r^{\frac{11}{2}}\,
    {\sqrt{1 - \frac{{\sin (\theta )}^2\,{r_*}}{r}}}}
\,
    \left( 15\,r^2 + 2\,r\,{r_*} + 
      6\,\cos (4\,\theta )\,{r_*}^2 - 
\right.\\
&&
\qquad\qquad 
\left. 
      6\,\cos (2\,\theta )\,{r_*}\,
       \left( -5\,r + {r_*} \right)  + 
      6\,{\sqrt{2}}\,\cos (3\,\theta )\,
       {r_*}^{\frac{3}{2}}\,
       {\sqrt{2\,r - {r_*} + \cos (2\,\theta )\,{r_*}}} \
\right)
\end{eqnarray}

\end{document}